\documentclass[a4paper,fleqn,usenatbib]{mnras}
\pdfoutput=1

\usepackage{mathptmx}

\usepackage[T1]{fontenc}
\usepackage{ae,aecompl}

\usepackage{graphicx}	
\usepackage{amsmath}	
\usepackage{amssymb}	

\newcommand{\ijimw}{Int.\ J.\ Infrared \& Millimeter Waves}
\newcommand{\aspconf}{ASP Conf. Ser.}

\newcommand{\KAPPA}{\textsc{kappa}}
\newcommand{\specx}{\textsc{specx}}
\newcommand{\cupid}{\textsc{cupid}}
\newcommand{\smurf}{\textsc{smurf}}

\newcommand{\makecube}{\textsc{makecube}}
\newcommand{\unmakecube}{\textsc{unmakecube}}
\newcommand{\mfittrend}{\textsc{mfittrend}}
\newcommand{\gsdacsis}{\textsc{gsd{\footnotesize{2}}acsis}}
\newcommand{\findback}{\textsc{findback}}
\newcommand{\findclumps}{\textsc{findclumps}}
\newcommand{\fixsteps}{\textsc{fixsteps}}

\newcommand{\ascl}[1]{\href{http://www.ascl.net/#1}{ascl:#1}}

\title[JCMT Heterodyne Data Reduction Pipeline]{Automated reduction of submillimetre single-dish heterodyne
  data from the James Clerk Maxwell Telescope using ORAC-DR}

\author[T. Jenness et al.]
{Tim~Jenness,$^{1,2}$\thanks{E-mail: tjenness@lsst.org (TJ)}\thanks{Present address: LSST Project Office, 933 N.\ Cherry Ave, Tucson, AZ~85721, USA}
Malcolm~J.~Currie,$^1$\thanks{Present address: RAL Space, STFC Rutherford Appleton Laboratory, Harwell Oxford, Didcot, Oxfordshire OX11~0QX, UK}
Remo~P.~J.~Tilanus,$^1$\thanks{Present address: Leiden Observatory, PO Box 9513, 2300 RA Leiden, The Netherlands}
Brad~Cavanagh,$^1$
\newauthor
David~S.~Berry,$^1$\thanks{Present address: East Asian Observatory, 660 N.\ A`oh\=ok\=u Place, Hilo, HI~96720, USA}
Jamie~Leech,$^1$\thanks{Present address: Department of Physics, University of
  Oxford, Denys Wilkinson Building, Keble Road, Oxford, OX1 3RH, UK}
and
Luca~Rizzi$^1$\thanks{Present address: W.\ M.\ Keck Observatory, 65-1120 Mamalahoa Hwy, Kamuela, HI~96743, USA}\\
$^1$Joint Astronomy Centre, 660 N.\ A`oh\=ok\=u Place, Hilo, HI~96720, USA \\
$^2$Department of Astronomy, Cornell University, Ithaca, NY~14853, USA
}

\date{in development; in original form 2014 December 2}
\pubyear{2015}

\begin{document}
\label{firstpage}
\pagerange{\pageref{firstpage}--\pageref{lastpage}}
\maketitle

\begin{abstract}

  With the advent of modern multi-detector heterodyne instruments that
  can result in observations generating thousands of spectra per
  minute it is no longer feasible to reduce these data as individual
  spectra. We describe the automated data reduction procedure used to
  generate baselined data cubes from heterodyne data obtained at the
  James Clerk Maxwell Telescope. The system can automatically detect baseline regions in
  spectra and automatically determine regridding parameters, all
  without input from a user. Additionally it can detect and remove
  spectra suffering from transient interference effects or anomalous
  baselines. The pipeline is written as a set of recipes using the
  ORAC-DR pipeline environment with the algorithmic code using
  Starlink software packages and infrastructure.  The algorithms
  presented here can be applied to other heterodyne array instruments
  and have been applied to data from historical JCMT heterodyne
  instrumentation.

\end{abstract}

\begin{keywords}
submillimetre: general --
methods: data analysis --
techniques: image processing --
techniques: spectroscopic

\end{keywords}

\section{Introduction}
\label{sec:intro}

As heterodyne receivers have progressed from single-detector
instruments
\citep{1992IJIMW..13.1487P,1992IJIMW..13..647D,1992IJIMW..13.1827C} to
small focal-plane arrays
\citep{2003SPIE.4855..322G,2004A&A...423.1171S} to 16-element arrays
such as HARP at JCMT \citep{2009MNRAS.399.1026B}, and beyond
\citep{2012SPIE.8452E..04K,2014SPIE.9153E..27H}, and correlators have improved such that
we can easily obtain spectra at 10\,Hz with 8192 channels, data rates
have increased substantially such that it is now common-place to take
a short observation resulting in thousands of spectra. This is only
going to become more challenging with the advent of instruments with
64,000 channels and dual-waveband arrays each of which consist of 128
detectors, such as the CHAI instrument proposed for CCAT
\citep{2014SPIE9152-109}
or KAPPa successors \citep{2014SPIE.9153E..0KW}.

The work described in this paper follows the installation of the ACSIS
(Auto-Correlation Spectral Imaging System) digital autocorrelation
spectrometer at the JCMT \citep{2009MNRAS.399.1026B}. ACSIS was
developed to provide a state-of-the-art spectroscopic backend for the
then forthcoming 16-element SIS mixer-based HARP focal-plane array \citep{2003SPIE.4855..338S}. As
well as having the capability to deal with 16 receptors at once, the
ACSIS correlator was designed to be capable of delivering new wideband
(up to 2\,GHz) and high-resolution (down to 30\,kHz) observing
modes. The ACSIS spectrometer was initially commissioned with the
existing single mixer instruments at the JCMT operating at 230\,GHz,
350\,GHz, 470\,GHz and 690\,GHz. While this work was being carried
out, the Telescope Control Systems (TCS) and Real Time Sequencer (RTS)
control systems \citep{2002SPIE.4848..283R} for the JCMT were
rewritten in preparation for HARP/ACSIS and SCUBA-2
\citep{2013MNRAS.430.2513H} receivers. ACSIS,
together with the improvements in telescope software infrastructure,
offered many observing advantages over the previous single
receptor-capable spectrometer, the DAS \citep[Dutch Autocorrelation
Spectrometer;][]{1986SPIE..598..134B}. Importantly, these included
support for a much wider variety of telescope observing modes, which
could be processed by the reconfigurable real-time online data reduction
system \citep{2000ASPC..216..502L}. The HARP/ACSIS
science goals involved mapping the intensity and characterising the
dynamics of cold molecular species in the ISM (such as CO and HCN), as
part of a programme of large scale JCMT legacy surveys
\citep{2010HiA....15..797C}. These legacy surveys involved wide area
mapping of both molecular components (using HARP/ACSIS) and dust
components using SCUBA-2 both within the
Galaxy \citep{2007PASP..119..855W} and for external galaxies
\citep{2009ApJ...693.1736W}.

In submillimetre astronomy, data reduction packages such as
\htmladdnormallinkfoot{\textsc{class}}{http://www.iram.fr/IRAMFR/GILDAS}
\citep[][\ascl{1305.010}]{2005sf2a.conf..721P} and
\specx\ \citep[][\ascl{1310.008}]{SPECX,1990JCMTP...9...25P} were developed that worked well with
single-detector instruments. Scripting interfaces and tools for
curating collections of spectra were insufficient as the data rates
increased and data pipelines \citep[e.g.,][]{1995ASPC...75..117W} and
algorithms that work on the full data set
\citep[e.g.,][]{2002ASPC..278..329M} were suggested. The ACSIS online data
reduction system \citep{2000ASPC..216..502L,2000SPIE.4015..114H},
delivered to the JCMT in 2005, aimed to deal with the data-rate issues
by providing a real-time pipeline that co-added the calibrated
spectra, with optional baselining, into a data cube with two spatial axes and one spectral axis.  This strategy
was forced on us given the computer resources available when ACSIS was
being designed and developed and was known to have risks associated
with it. Coadding spectra into the cube allowed for impressive ``data
compression'' for stare and jiggle observing modes, which repeatedly
observe a fixed set of positions, but the gains were less in scanning
observing modes.
The gridded, baselined and co-added data cube was the product that was archived
and taken away by the astronomer for further analysis, although it was
also possible to store the raw data in CASA (\ascl{1107.013}) measurement sets
\citep{2012ASPC..461..849P}\footnote{At the time this was being
developed CASA was known as AIPS++ \citep{2004ASPC..314..468M}},

There are obvious downsides associated with this approach. The
observing system required that the cube parameters be specified and
pre-selected by the observer in the Observing Tool
\citep{2002ASPC..281..453F}. It was also
necessary that the observer specify the baseline regions and any
frequency binning required. Since such a process as a whole is
irreversible, on a fundamental level it is not well-adapted to
astronomical research where observations are not well characterized
and pre-set values may turn out to be a poor choice. It is also not
compatible with modern approaches to flexible scheduling
\citep{2002ASPC..281..488E} where the astronomer planning the
observations is not doing the observing. Consequently,
when it became clear in 2006 that computers were fast enough and disk capacity
large enough to be able to store the observed spectra without the need
of real-time regridding and coadding, this further data reduction was
migrated to a separate loosely coupled pipeline system and the calibrated
spectra became the raw data written by the instrument and archived by
the observatory.

A post-acquisition data reduction pipeline has clear advantages and is
now implemented in some form at most modern observatories. Foremost,
the process can be repeated both to correct for errors, but also, if
necessary, to iteratively fine-tune the reduction to the
characteristics of the individual observations. Although tunable, the
scripts or recipes that drive the pipeline impose a standard of
reduction that can include advanced techniques and sophisticated
quality-assessment checks that would be hard for the average user to
master. Furthermore different incarnations of the pipeline can be
deployed in different environments: a basic version to provide
near-time feedback at the telescope during observing, a comprehensive
version at the observer's home institution for the advanced reduction,
and a version at an archive center that can process the result of a
user query, possibly retrieving and combining observations from
different projects.

\section{Heterodyne Data Reduction Pipeline}

The pipeline at the JCMT is implemented by writing heterodyne data
reduction recipes for the ORAC-DR pipeline infrastructure
\citep[][\ascl{1310.001}]{2011tfa..confE..42J,2015A&C.....9...40J}
that was already in use at the telescope with SCUBA
\citep[e.g.,][]{1999ASPC..172..171J}. These recipes are written in
Perl to simplify control flow but use Starlink applications \citep[see
e.g.,][]{2014ASPC..485..391C} for the per-pixel data processing. The
main Starlink applications used for the heterodyne pipeline are
\smurf\ \citep[][\ascl{1310.007}]{SUN258} for instrument-specific
algorithms, \cupid\ \citep[][\ascl{1311.007}]{2007ASPC..376..425B} for
determining emission regions, and \KAPPA\
\citep[][\ascl{1403.022}]{SUN95} for general-purpose data processing.

The first guiding principle for the overall
design is for the pipeline to deliver sensible results based only on
information in the data files themselves, without any further user
input. This driver for minimizing user input leads to two key requirements
for the pipeline: the parameters of the resultant data cube must be
derived solely from the positions of the individual data samples, and
the spectral baseline regions must be determined automatically by
looking at all the spectra together.
On a more advanced level, it requires for the pipeline to be
able to detect and remove bad spectra as well as carrying out quality
assurance (QA) tests (see \S\,\ref{sec:qa}). The latter are critical
for the JCMT Legacy Survey projects
\citep{2007PASP..119..855W,2009ApJ...693.1736W,2007PASP..119..102P}
who want to ensure they receive data of consistent quality.  Quality
assurance tests not only need to enable the judgement of the data against absolute
criteria, they also need to test the self-consistency of the overall
data set being processed, which possibly attempts to combine
observations taken under vastly different conditions and even
different instrument configurations.

A second design principle for the pipeline is that its products
retain the spatial and spectral resolution of the original data.  The
data reduction relies heavily on spatial and spectral smoothing of the
data cubes to improve S/N and isolate the three-dimensional nature of astronomical
objects. Results from such an analysis are used to mask the original
data, thus maintaining the original resolution. An example is
baselining: the data cube is smoothed to a lower resolution,
both spatially and spectrally, in order to auto-detect emission-free
baseline regions. The result is then used to mask the original
un-smoothed data cube and perform the actual fit of the baselines.

A further design principle is for the pipeline to be iterative: the
final results can be used to refine the reduction of the individual data
sets for which the S/N may be much worse. These in turn are then used
to re-derive the results. Although the pipeline in principle can be
configured with an arbitrary number of loops, in practice only a
two-step process is needed. For scanning observations implementing
this iterative process some of the results, such as baseline masks,
are required to be re-expanded in the time domain. Further optimizations are optional in
the second step. For example, while the baseline fit is linear
during the first step, with secure baseline regions higher order fits
can be used during the second step.

It is not possible to run a general heterodyne pipeline without any a
priori user-input at all. The reason for this is that at a fundamental
level it is impossible to distinguish between, for example, a broad spectral
line and a non-linear baseline feature, or a narrow emission feature
and a spurious data spike extending over a few channels. For JCMT data
different pipeline recipes have been designed optimized towards,
for example, the broad and narrow-line case (see \S\ref{sec:alt}). The observer specifies the choice
of pipeline reduction recipe in the Observing Tool when preparing the
observations. The desired recipe is documented in the meta-data of
each data file. Recipes will be discussed in more detail below.

The JCMT heterodyne pipeline has two operating modes. The default
behaviour is for the pipeline to generate the best possible data
products without regard to efficiency. This is generally what is
required by scientists at their institutions and the mode the pipeline
is run in from within the JCMT Science Archive
\citep[JSA;][]{2015Economou,2008ASPC..394..565J}. The other mode is a
cut-down version of the recipes that runs at the JCMT itself during
observing. This pipeline has constrained timing requirements and can
not perform many of the advanced processing features provided by the
main pipeline. Its role is to provide simple quality-assurance
information and basic coadds to the observer and it will not be
discussed further in this paper.

\subsection{Observing modes}
\label{sec:obsmodes}

Single-dish heterodyne submillimetre observing involves making
observations through an atmosphere which can have a significant, time
varying opacity which depends on the integrated precipitable water
vapour above the observing site. In order to reduce the effects of
this time-varying atmospheric emission, several heterodyne observing
modes are typically used at the JCMT. Position switching involves
moving the telescope to observe through the same patch of atmosphere
next to the astronomical object. Beam switching involves chopping the
movable secondary mirror of the JCMT at a rate faster than the
typical time variation of the atmospheric emission (a few
Hz). Finally, frequency switching involves rapidly retuning the
receiver's local oscillators to shift the position of the spectral
lines in the intermediate-frequency (IF) passband. None of these methods are completely
effective, especially in poorer weather conditions.  Imperfect
atmospheric removal can lead to residual unwanted baselines, which may
have linear, polynomial or sinusoidal forms. In addition, the effects
of standing waves, cable flexure and unwanted frequency dependent
phase slopes within the IF system itself can also lead to unwanted
baseline features. These can be particularly problematic, in that
they are often poorly described by simple linear or low-order polynomial
fitting and they take the form of slowly varying sinusoids or
rapidly varying ``wiggles'' (see \S\ref{sec:non-linearity}).
Characterising and, if possible, removing these baselines in a fashion
which is as highly automated as possible is an important design
requirement of any potential data reduction scheme.

Full details of the JCMT heterodyne observing modes can be found in
\citet{2009MNRAS.399.1026B}, but we provide a summary here. The most
common observing mode is the scan mode used to map large area. In this
mode the telescope fills a rectangular area by scanning a
boustrophedon pattern: first scanning in one direction, then moving up
to the next row and then scanning in the reverse direction. The
K-mirror rotates such that the array is angled relative to the scan,
allowing a fully-sampled patch of sky to be observed in a single
pass. Subsequent observations can repeat the area with the scan
direction perpendicular to the first to ensure that each position on
the sky is measured by different detectors and to help minimize
``print through'' of the scan pattern.

The Jiggle mode is used for areas the same size as the HARP field of
view. Here the secondary mirror moves to fill in the gaps between the
array elements while the telescope tracks the target position.  While
jiggling the secondary typically is fast and efficient, given that
adjacent pixels in each area of the map tend to be measured by a
single detector, this mode is non-optimal in the case where detector
performance and stability significantly differ across the array. In
such cases high quality maps require many repeats with, for example,
the K-mirror at various angles and with the telescope moved to several
offsets to ensure that different detectors contribute to each point in
the map. For this reason, unless the source of interest is
sufficiently compact compared with the field of view, for HARP it is
usually better to do a small scan map for high-fidelity imaging.

\subsection{Heterodyne Data Files}
\label{sec:format}

Spectra from HARP arrive asynchronously from data-acquisition paths
that in parallel handle data from the individual detectors and long
observations are split over several files. One of the basic first data
reduction steps is thus collating and ordering of spectra into a time
and detector sequence.  A description of the heterodyne raw data file
format used at the JCMT can be found in Appendix~\ref{sec:rawdata}.

For scanning observations different detectors will observe almost
the same sky position and will contribute to the same output pixel in
the final gridded data-cube. The data reduction will thus need to keep
accurate track of the performance of each detector and the noise of
each of the spectra individually. Moreover, since different detectors
and varying channels may intermittently have problems, such as
interference, flagging and noise-tracking may need to be propagated on a
per-channel basis. Heterodyne data cubes at the JCMT for this reason
have a variance array and flagging bits associated with each data point.
While this roughly doubles the data volume, this is the only way to
ensure correct error statistics and assignment of relative weights
when data are combined.

\begin{figure}
\includegraphics[width=\columnwidth]{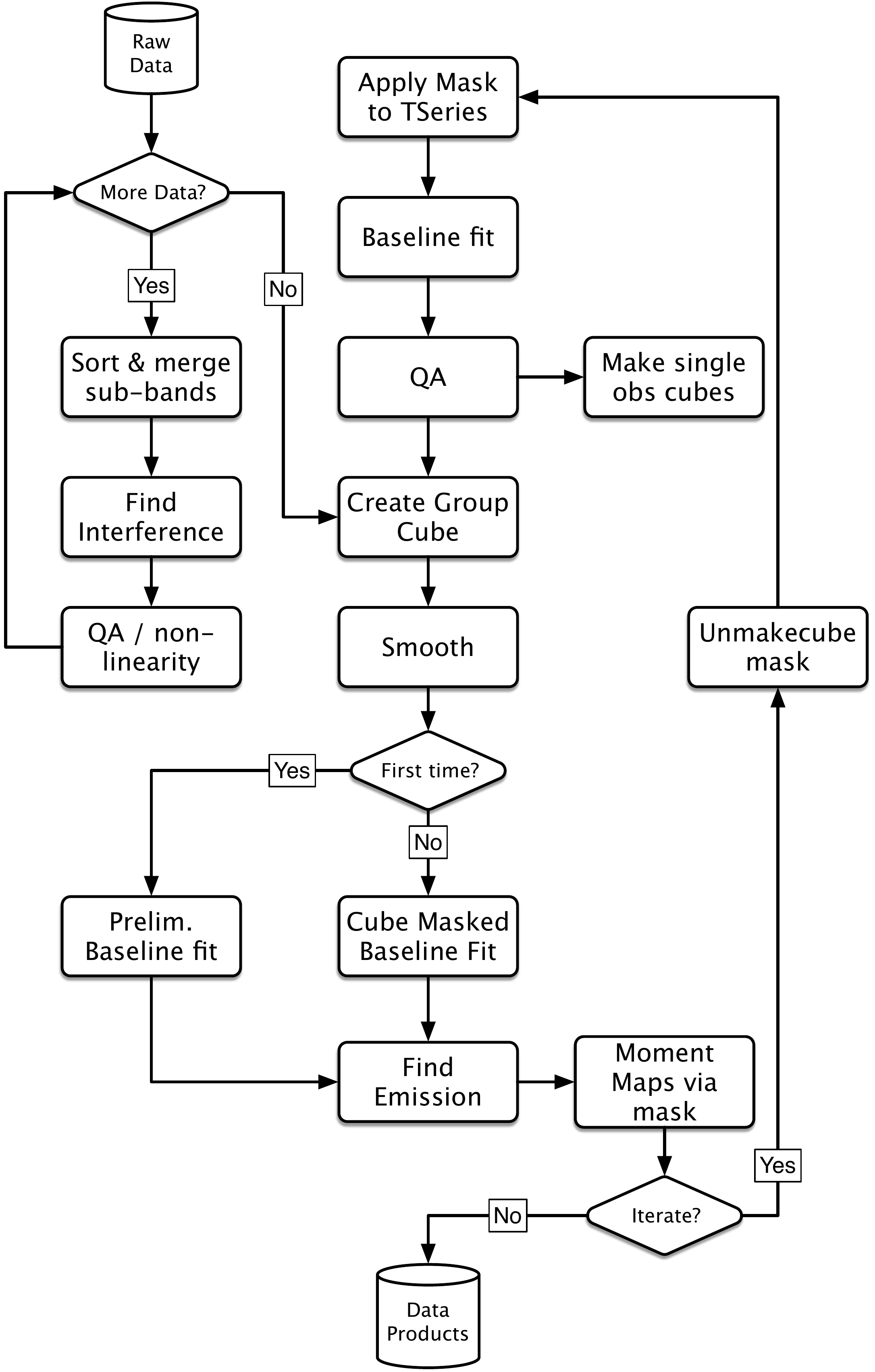}
\caption{Flow chart of the pipeline recipe, including the initial step
  where individual observations are analysed. \emph{QA} here indicates
  a quality-assurance test (see \S\,\ref{sec:qa}).
}
\label{fig:flowchart}
\end{figure}

\section{Pipeline processing}

Fig.~\ref{fig:flowchart} shows the key steps in the automatic
pipeline reduction of heterodyne data at the JCMT. An initial phase
checks individual observations and prepares them for coadding and
gridding into the \texttt{group} cube (see the upper left of the figure).
This is followed by a group
phase where baselines are removed and emission selected in the coadded
cube.  Next follows an iterative step: the resulting masks are
re-expanded into the time-domain and applied to the individual
observations with the aim at producing a better group file. Given that
accurate baseline (emission-free) regions are available, higher
order baselines can be removed from the individual observations and
tighter QA tests can be applied. These three phases are discussed in
more detail next.

\subsection{Preprocessing the individual observations}

The initial phase works on the ungridded individual observations,
which in essence are a time series of spectra from each individual
detector with a typical dump time of 1 second or faster in case of scan
observations.  Not all recipes perform exactly the same reduction, but
common steps for each observation are:
\begin{itemize}
\item Combine all data files belonging to the observation
\item Sort spectra by time and detector
\item Remove median signal per time step (i.e.\ common-mode signal over
all detectors and channels)
\item Basic despiking and interference flagging
\item Combination of overlapping spectral windows (sub-band merging)
\item Basic QA assessment checks: flag noisy detectors, unexpectedly noisy
observations, spectra with large noise gradients etc.
\item (Optional) Remove basic linear baseline and grid the individual
observation.
\end{itemize}

The preprocessing aims at preparing individual observations for
coadding and removing obviously problematic spectra and channels.
Note that, although gridded cubes can be produced for manual
inspection of the individual observations, this is not needed for
further processing because the coadded group cube is produced by
combining the ungridded data from all observations rather than
averaging gridded cubes.  This avoids having to resample already
gridded cubes onto a common frame and allows also for the combination
of mosaicked or even separate fields into a single cube.

\subsection{Group processing}

After the preprocessing is completed a gridded group cube is created
from all the observations and processed further. Here is where the
three-dimensional structure of the data becomes critical for the analysis
and the different data reduction recipes become more specific. The first
distinction is whether single spectral lines extend over a significant
fraction of the available band or not. If not, the second distinction
is whether smoothing should be spatially biased (e.g., narrow spectral
lines) or spectrally biased (broader lines and/or strong velocity
gradients within the field-of-view). The smoothing is done by applying
a tophat convolution in three dimensions ($x$, $y$, $v$) with a smoothing
factor in each direction that depends on the recipe.  By default a
recipe biased towards spatial smoothing will have smoothing factors of
($x=5$, $y=5$, $v=10$) pixels\footnote{For scan maps the pixel sizes
  are approximately half beam whereas for jiggle maps they are either
  half or one-third of a beam. The spectral channels can vary in size
  from 0.05\,km/s to 0.4\,km/s depending on correlator setup.},
 whereas one biased towards spectral smoothing
will be ($x=3$, $y=3$, $v=25$). Note that both type of recipes will reduce
the noise per pixel by a factor of about 15. While the smoothing
factors in velocity may seem excessive, they reflect the very high
resolution and frequency coverage of present-day correlators
compared with typical width of spectral lines.

Smoothing the data will emphasize extended regions of emission in the
three-dimensional data cube, but bias against weak narrow-line features that are also
point-like. Projects that search for such objects need to use specialized
software to analyze the basic, unsmoothed, group file directly.

For standard processing, the resulting smoothed high-S/N cube
is then used by the baseline-fitting routine, \mfittrend\ from
\KAPPA\  (see \S\ref{sec:mfittrend}),
to determine emission-free spectral windows.
The details of this process depend on the recipe: ranging from using two
windows with a set width at each end of the frequency band
for the ``broad-line'' case, to a full automatic search for emission-free
windows in case of multi-line spectra. Note that using the smoothed data has
two advantages. Firstly, the significantly higher S/N allows for a better
distinction between emission and emission-free regions. Secondly, smoothing
also spreads extended emission regions somewhat, both spatially and spectrally,
resulting in a conservative estimate of emission-free regions that is better suited for
the fit of low-order baselines in this first pass. Once \mfittrend\ has
had this first pass at the baselines, a mask is written out containing the
baseline regions of the smoothed data cube. This mask is applied to the original
unsmoothed data cube resulting in a cube consisting solely of
baselines. \mfittrend\ is then used again but this time fitting
the entire (masked) spectrum, fitting a linear (or, optionally, higher-order) baseline
to each spectrum in turn. The baselines are then finally subtracted from the
original data cube.

With proper baselines subtracted the data cube can be analysed using more
holistic approaches to isolate emission associated with the astronomical
target. This step is critical since for most target fields the number of pixels with
noise far exceed those with a signal rendering a simple collapse along an axis
or moments analysis of the baselined data cube virtually unusable. Instead
the JCMT pipeline uses a clump-finding algorithm to isolate and identify
emission features.
The Starlink \cupid\ application contains
a number of clump-finding algorithms that work in three dimensions.
The choice of a particular algorithm is not critical and either FellWalker \citep{2015FW} or
Clumpfind \citep[][\ascl{1107.014}]{1994ApJ...428..693W} can be used.
As for the baseline fit, the clump find is performed on a smoothed version
of the baselined group cube, resulting in masks of emission regions that
are applied to the unsmoothed baselined data. A moments analysis routine
(see \mbox{\S \ref{sec:moment}}) can then be used with the masked data set to
extract, for example, a total emission map or velocity field. Since higher-order moments are
more sensitive to noise features, different S/N cutoffs are used for the
clump masks used for the different moments. This approach results in deep
total emission (zero-moment) maps as well as reliable anomaly-free
velocity (first-moment) maps.

\subsection{Iterative processing}

The group processing described above delivers a baselined data cube,
baseline and clump masks, and moments maps. Iterative processing uses
the baseline mask from the group data cube to improve the baseline fit
of the individual observations. These in turn are used to generate an improved
group file, which is then reduced using the same steps as in the first iteration.

In order to apply the masks to the raw data, the masks need to be resampled
in a time and detector domain. This is done using the \smurf\ \unmakecube\ task.
This application
does the opposite of \makecube\ -- whereas \makecube\
generates a gridded cube from a time-series cube, \unmakecube\ generates
a time-series cube from a gridded cube, using a supplied time-series cube
as a template to define the spatial and spectral positions at which the
gridded cube is to be sampled. In this way \unmakecube\ is used to generate
a time-series cube from the group data cube mask. This time-series cube can
then be used to mask the original time-series allowing the baseline to be
fitted to each individual input spectrum.

This is critically important for generating properly baseline-subtracted
cubes of the individual observations but can also
be important for quality-assurance tests. The enhanced baseline
subtraction can ``resurrect'' spectra that were originally determined
to be of poor quality and not used in the final cube. This iterative
cube production with enhanced QA can lead to minor improvements in
quality of the final product.

A main objective for the iterative step was to allow for non-linear baseline fits
(except for the broad-line recipe), both for individual observations as well as
the group cube, given that accurate baseline windows have been determined.
Initially the pipeline was configured to use up to fifth-order baselines in
the second iteration step. In general this was very effective in successfully
removing even high-order baselines from observations taken when the
conditions or instrumentation was unstable, without negatively impacting
the vast majority of the observations for which linear or second order baselines
were sufficient.  However, a subset of users objected to an automatic fitting of
non-linear baselines and the default pipeline behaviour was changed to
fit linear baselines only, leaving higher-order fitting to a custom pipeline reduction
by the users themselves.  In our opinion this is unfortunate since it significantly
diminishes the benefits of the iterative scheme when, for example, pipeline processing
in place for the JSA.

\subsection{Customization}

The JCMT heterodyne pipeline is highly customizable. At the highest level, recipes
are simple text files with calls to ``primitives''. Each primitive has a list of
parameters associated with it that can be set or changed by editing the recipe.
The pipeline can then be instructed to use this custom recipe instead of the default one.

The parameters associated with the primitives, however, give access to only a small
number of the full set of parameters allowable for the Starlink routines. A user can access
a subset of these by setting up a special configuration file called a recipe
parameter file. A single configuration file for a project can
be used to assign different parameter values for observations of different fields and
different observing frequencies. An example is setting an allowable velocity range for
spectral features: the default pipeline makes no assumption about this which often results
in including in its analysis large sections of the data cube with only noise. This can lead
to problems from the cumulative effect of 4- or 5-sigma noise spikes. Pre-specifying the
allowed velocity range typically is an effective way to improve
pipeline results. Recipe parameters can also be used to control how to
bin up the frequency scale, specifying the output grid and any
regridding parameters, and also whether to enable or disable
flatfielding (\S\,\ref{sec:flat}) and bad-baseline filtering (\S\,\ref{sec:badbase}).

The JSA is configured to accept ``user'' defined configuration files: if such
a file exists for the observation being requested, it will be used in place of the default
pipeline reduction of the data. The pipeline is designed at the moment
to always start from the raw data and can not begin part way through a
recipe; modifying a recipe parameter that is only used late in the
processing still requires that all the initial processing is performed.

\subsection{Improvements}

Although the JCMT heterodyne pipeline has proven to be effective in delivering
high-quality results, there are a number of potential improvements that remain
un-implemented or unexplored due to lack of resources.

\begin{itemize}
\item  Higher-order baseline fits during iterative processing. As discussed, the
iterative processing aspect of the pipeline aimed at allowing for higher order
baseline fits during subsequent iterations, but a subset of users objected
to this. This can be addressed by a routine that critically examines the
parameters associated with fits to either flag high-order baselines
or to optimize the choice of order used. The fraction of poor fits associated
with a particular fit order can be used as quality-assessment parameter.
\item Fields with narrow and broad profile features.  The current pipeline
is ill-equipped to deal with fields that mix narrow and broad spectral
lines, such as found in the central regions of galaxies or compact
outflow sources. Given that the regular baseline routine also produces a
cube with ``noise-free'' fitted baselines, this cube can be analyzed similar to a
broad-line observation to extract and check for broad
spectral components that were erroneously subtracted.
\item Similar tactics could be applied in reducing continuum observations:
in the current default continuum emission recipe no baseline is subtracted. Instead the continuum
level could be determined from the median level of each baseline fitted.
\item Adding a routine that ``auto-detects'' the main velocity range based
on a statistical analysis of the distribution of baseline windows or detected clumps
within the cube. As discussed in the previous section, restricting the allowable
velocity range is a simple way to improve the pipeline result. In addition,
such routine can flag serendipitous sources or spurious features outside
a primary velocity range.
\item Using a profile fitting routine instead of a moments analysis. Recently a
comprehensive Gauss-Hermite multi-profile fitting task \textsc{Fit{\footnotesize{1}}d}
was added to the \smurf\ package that produces a cube with the fitted profiles as well
as cubes with the fitted parameters (amplitude, position, width, etc.) for
each spectral feature. Gauss-Hermite functions can fit complex, asymmetric
profiles and for certain projects, such as a spectral survey, this may be more
appropriate than fitting moments.
\end{itemize}

\section{Component processes}

\subsection{Determining Cube Parameters}
\label{sec:makecube}

\begin{figure}
\includegraphics[width=\columnwidth]{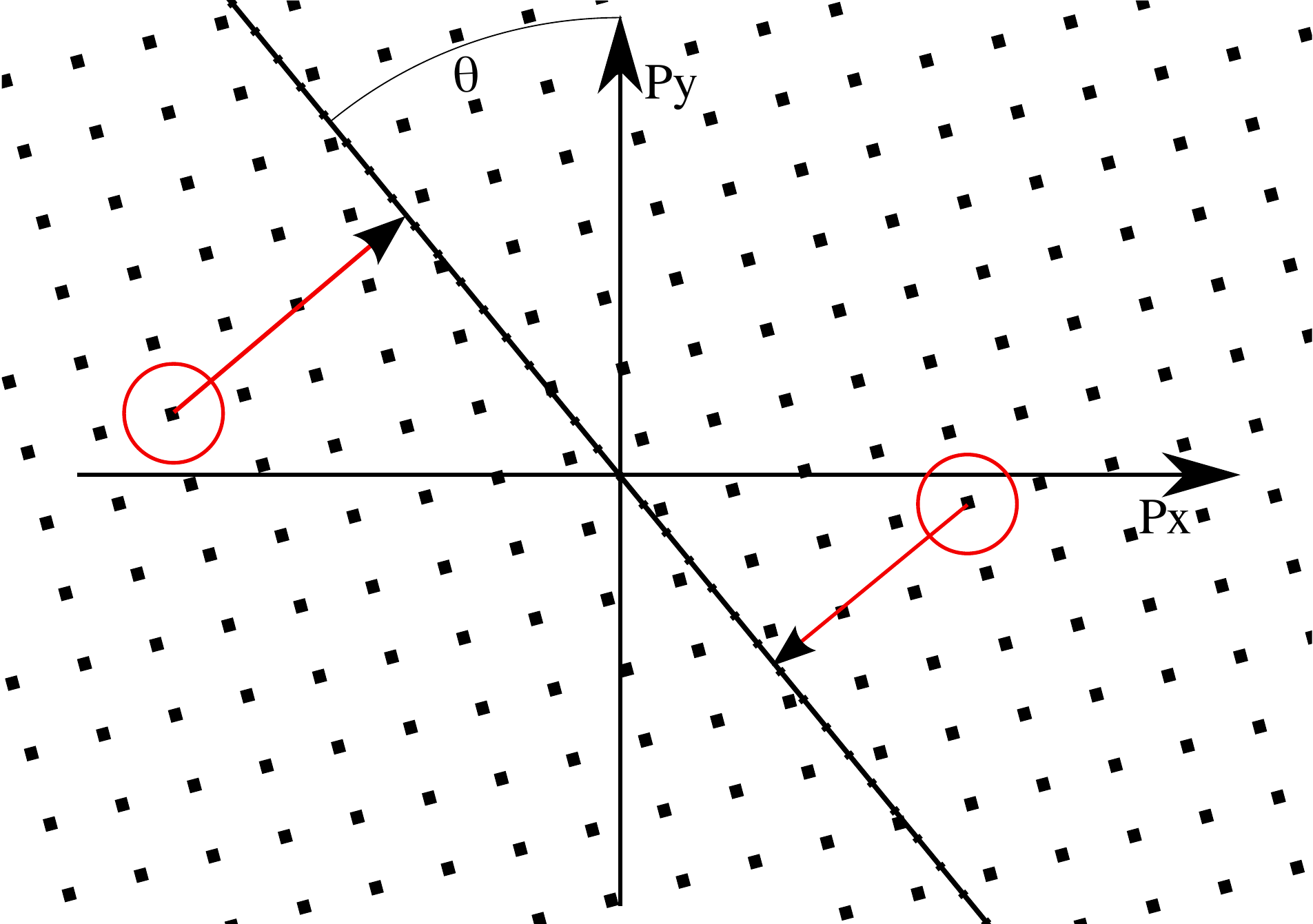}
\caption{The Autogrid algorithm works by projecting each spectrum
  position onto a straight line at an arbitrary angle $\theta$, and then
  forming a histogram of the number of samples at each point along
  this line. At the optimal value of $\theta$, the projected positions
  line up, giving strong periodicity in the histogram.}
\label{fig:autogrid}
\end{figure}

Since the output pixel grid need not be specified in advance, software
is required to determine the pixel grid from the data itself. In the
\smurf\ package data cubes are created from calibrated spectra using the
\makecube\ command. \makecube\ can be given an externally specified
grid but also has an \texttt{autogrid} option that leaves optimal grid
determination to the application itself.

To maximize overall aperture efficiency, detectors in HARP are spaced
30\,arcsec apart, which introduces a natural scale of order an arcminute
into the observations, even if scan and jiggle observations will be
sampled on much smaller scales than that. Autogrid first projects the
supplied sky positions into pixel positions using an arbitrary tangent
plane projection that has 1~arcmin square pixels with North upwards
and the target position (or the first supplied sky position if no
target position is available) at pixel (1,1).

It then projects each of these pixel positions onto a straight line
passing through pixel (1,1) at an angle, $\theta$, to North (see
Fig.~\ref{fig:autogrid}). This line is divided up into sections of
length 1~arcmin, and a histogram formed of the number of projected
positions that fall in each section. The amplitude and wavelength of
any periodicity in this histogram is found by looking at the
auto-correlation of the histogram (the amplitude is the
auto-correlation at zero shift, and the wavelength is the shift at the
first significant peak in the auto-correlation function).

This is repeated for many different line orientations in order to find
the value of $\theta$ (line orientation) that gives the strongest
periodicity in the histogram. This orientation is used as the
direction for the X pixel axis in the final grid. The corresponding
wavelength is used as the pixel spacing on the X axis. The wavelength
of the periodicity perpendicular to this direction is then found and
used as the pixel spacing on the Y axis.

Finally, the reference-pixel coordinate is shifted by up to one pixel
on each axis in order to minimise the sum of the squared distances
from each pixel projected sample position to the nearest pixel
centre.\footnote{If the positions do not form a regular grid, an option is
available to create a 1-dimensional list of spectra in which the
position of each spectrum is recorded explicitly in a table using
the FITS-WCS \texttt{-TAB} algorithm \citep{2006A&A...446..747G}.}

\subsection{Combining Spectra}
\label{sec:combine_spectra}

The JCMT heterodyne pipeline uses Starlink routines which
have been designed to maintain accurate variance and flagging data.
Nevertheless, this is not sufficient to fully deal with the issue of
bad data. A simple illustration is the coadding of two spectra with a
different DC level: the DC level of the result will be the average
level. However, if one of the spectra has a bad channel, adopting the
one remaining point in the output would result in a positive or
negative spike since its DC level will not be average. In other words,
a policy for dealing with bad data that is acceptable in one situation,
i.e.\ spectra without a DC level or corrected for the DC level, can be
problematic in a different situation.

In combining data, the gridding software recognizes three schemes for
dealing with bad data:
\begin{description}
\item[\texttt{AND}] An output pixel will be bad only if all the input
   pixels are bad. This scheme will produce the least number of bad
   output pixels, but memory requirements can be excessive and are much
   larger than for the other two schemes. It also is affected by issues such
   as the DC-level problem discussed above.
\item[\texttt{OR}] An output pixel will be bad if any of the input
   pixels are bad. This scheme will produce the most bad output pixels,
   but a more homogeneous noise across the image and avoids many
   of the issues associated with the AND scheme.
\item[\texttt{FIRST}] Only spectra that have the same bad pixel mask
   as the first spectrum contribute to the output spectrum. It
   produces fewer bad pixels in the output than the \texttt{OR}
   scheme without the large memory footprint of the \texttt{AND} scheme.
   This scheme is useful in the absence of intermittent problems and where
   the bad pixel mask, for example, results from a long-term instrumental effect, but
   comes with the risk of rejecting large amounts of otherwise good data.
\end{description}

The pipeline processing defaults to using the \texttt{AND} scheme but
can be overridden if either speed or memory is an issue.

\subsection{Cube forming}

Once the grid has been determined, the output spectrum at each
position is formed by averaging the nearby input spectra. Various
averaging schemes are available, the simplest being to place each
input spectrum entirely into the nearest output pixel. Other schemes
allow a two-dimensional kernel to be used to spread each input spectrum
out over a range of output pixels. Available kernels are those
supported by the AST library \citep{SUN211,2012ASPC..461..825B} and
include a simple bi-linear division between the four nearest
neighbours, a Gaussian kernel, and various flavours of kernels based
on a sinc function.

The resampling can use any of the three schemes described in
\S\ref{sec:combine_spectra} to determine how bad pixels are propagated
from input spectra into the output cube.

One complication is that the data cube for a large area survey is
potentially extremely large and many software packages do not support
data arrays with more than $2^{31}$ pixels\footnote{Much of the code
is written in Fortran using a signed \texttt{INTEGER*4}.}. We overcome
this by supporting the ability for the output cube from \makecube\ to
be split into tiles. These tiles share projection parameters and can
be recombined without further resampling if required. For
pixel-spreading techniques that are susceptible to edge effects care is
taken to ensure that sufficient border is included in the tiles such
that the values in the output would be identical to those resulting
from a single output data cube. The border region is flagged in an
associated bit mask to ensure that it can be disabled during further
mosaicking or combination steps as the data in the border region will
have edge effects and is duplicating data found in other tiles.

Heterodyne arrays currently require that the detectors have spacing
much larger than the beam such that the data are inherently
under-sampled. Image rotators and clever scanning modes can overcome
this deficiency to a certain extent but in many cases the spatial
distribution of samples is inherently uneven. There may be benefits
associated with using an unbiased linear interpolation method such as
\emph{Kriging} \citep[e.g.,][]{1990Cressie} rather than a simple
convolution kernel.

\subsection{Sub-band merging}

\begin{figure*}
\includegraphics[width=\textwidth]{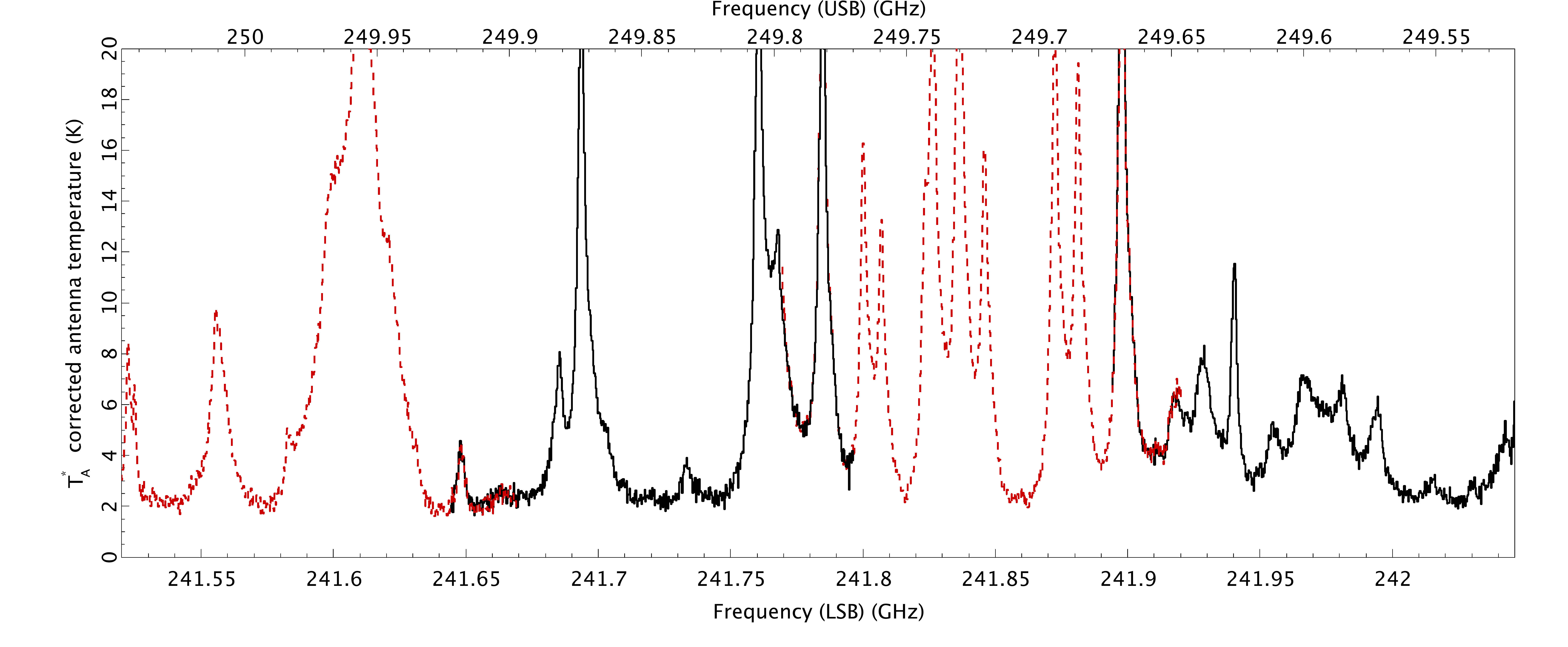}
\caption{A hybrid spectrum from an observation in Orion of multiple
  methanol transitions. The frequency scale is for the kinematic local
  standard of rest and the observations were taken with a rest
  frequency of 241.791\,GHz. These data were taken on 1998 December 20
  as part of project M98BA3I. The noisiest 15 channels have been
  removed from each end of the sub-bands spectral range for
  clarity. They would be removed as part of the merging process.}
\label{fig:hybrid}
\end{figure*}

The ACSIS IF system contains two local oscillators which perform two
stages of down-conversion, first to a \emph{parking} band between
1--2\,GHz and then to a 0--1\,GHz \emph{baseband} which is then sampled
by a 3-level ADC \citep{2000SPIE.4015..114H}. The down-converted bands
can be either 250\,MHz or 1\,GHz wide, and several of these
overlapping \emph{sub-bands} can be arranged in various ways to
achieve the required total IF frequency coverages. Typically the
sub-bands are arranged, by suitable LO tuning, to have areas of
overlap in frequency space, and these must be combined in software in
a process known as sub-band merging. The correlator is usually
configured such that the individual spectra overlap and also have
channels that are aligned to within a few per cent of a pixel.

The data acquisition system does not combine the sub-bands and so this
must be done by the pipeline. If the pipeline determines that it is
dealing with a hybrid mode observation the merging is done in
bulk. First the spectra are sorted by time (the ACSIS acquisition
computer does not guarantee that spectra will be written to files in
time order), then the overlap region is determined and the noisy ends
are trimmed before they are combined. The spectra can optionally have
their DC level adjusted before combining.

Fig.\ \ref{fig:hybrid} shows an example hybrid spectrum consisting of
four overlapping sub-bands from observations of methanol in Orion. In this
example correcting for any DC offset is complicated by the lack of
baseline region.

\subsection{Automated Baseline Removal}
\label{sec:mfittrend}

The Starlink \KAPPA\ task \mfittrend\ is used to calculate a
first-order baseline fit for each spectrum in the data cube
independently. The baseline region is estimated by a technique lent
from photographic surface photometry \citep{1998A&AS..127..367Y} but applied to one-dimensional
data.  A spectrum, or the mean spectrum over a region, is divided into
bins, typically 32.  Then a linear fit is made to the mean values and
outlier bins are excluded with progressive sigma-clipping leading to
an improved fit.  This rejection process generates a mask of deviant
bins, which is expanded back to elements in the original spectrum,
whose baseline is then fit without binning.

\mfittrend\ could attempt to refine the mask by narrowing the bin
widths within the rejected bins to pinpoint the emission and yield
more baseline to fit.  In practice the pipeline only determines a
first-order fit and the loss of a small fraction of the baseline
appears to make no significant difference.  Also it is better to be
conservative to ensure that no weak line velocity dispersion is
included to bias the fit's slope.  A further refinement is to perform
progressive sigma-clipping or use the histogram within each bin to
estimate the mode, thereby remove spikes and weak astronomical signal
from secondary lines that bias the baseline fit.

In the large majority of cases the masked spectrum will be free of
emission.  However, the method is not guaranteed.  One such case is if
the baseline is not linear, and such spectra are routinely rejected
(\mbox{\S \ref{sec:non-linearity}}).  More of an issue are very broad
lines that occupy a substantial fraction, more than half in some
cases, of the spectral range.

\subsection{Clumpfind and moments maps \label{sec:moment}}

The Starlink \cupid\ task \findclumps\ is used to finds clumps of
emission within each spectrum.
This is an essential step since moment maps can be compromised
if excessive baseline is included in the calculation. For example, in an
integrated-intensity calculation the inclusion of all the baseline noise can
hide a weak line.
Using a mask based on the detected clumps much improves
fidelity and improves upon using a simple threshold or
the smoothing scheme used in the \textsc{momnt} command in AIPS
\cite[][\ascl{9911.003}]{2003ASSL..285..109G}. As discussed in
\mbox{\S \ref{sec:makecube}}
observations of extended regions are potentially spread over multiple
data cube tiles, which must be processed independently, and the resulting
moment map is created by
mosaicking the individual submaps, taking into account the border
regions. This is configured such that no resampling is required as
\makecube\ ensures that all tiles are on the same pixel grid.

\begin{figure*}
\centering
\includegraphics[width=\textwidth]{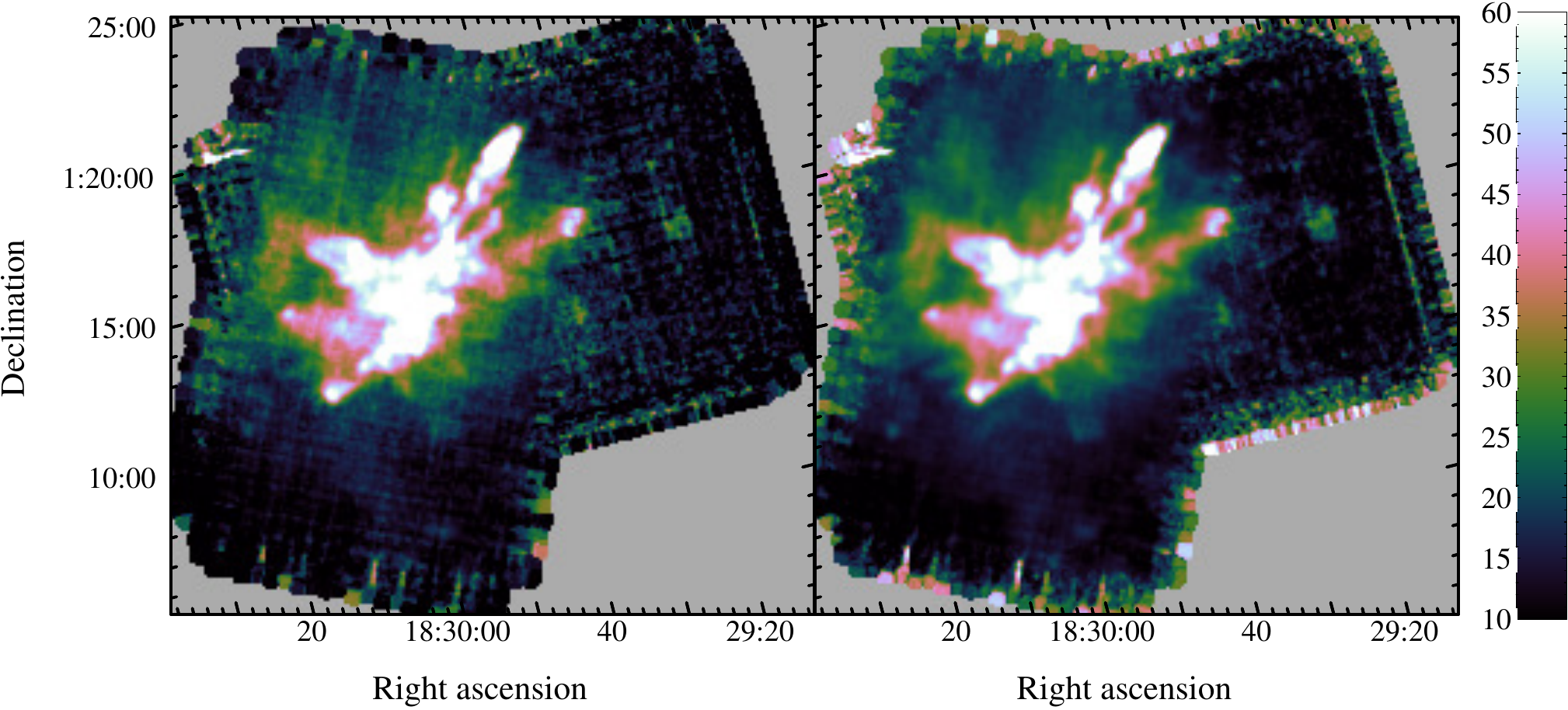}
\caption{Two integrated intensity images from the same set of
  observations of Serpens from 2007 using the same display
  parameters. The left figure uses a naive sum over a significant part
  of the baseline. The right figure uses the automated baseline
  masking. The key is in units of K\,km\,s$^{-1}$.}
\label{fig:integ}
\end{figure*}

Fig.\ \ref{fig:integ} compares integrated intensity images calculated
in two different ways from a single data cube generated by the
pipeline \citep[see][for details of earlier
reductions of these data]{2010MNRAS.409.1412G,2010A&A...523A..29D}. This data set
has some interference in a few spectral channels of a few detectors,
which has not yet been handled by the main pipeline
processing. Nevertheless, the right-hand image shows no sign of the grid
printing through and shows much more dynamic range than the naive
integrated intensity image.

\subsection{Removal of Bad-Baseline Spectra}
\label{sec:badbase}

\begin{figure}
\includegraphics[width=\columnwidth]{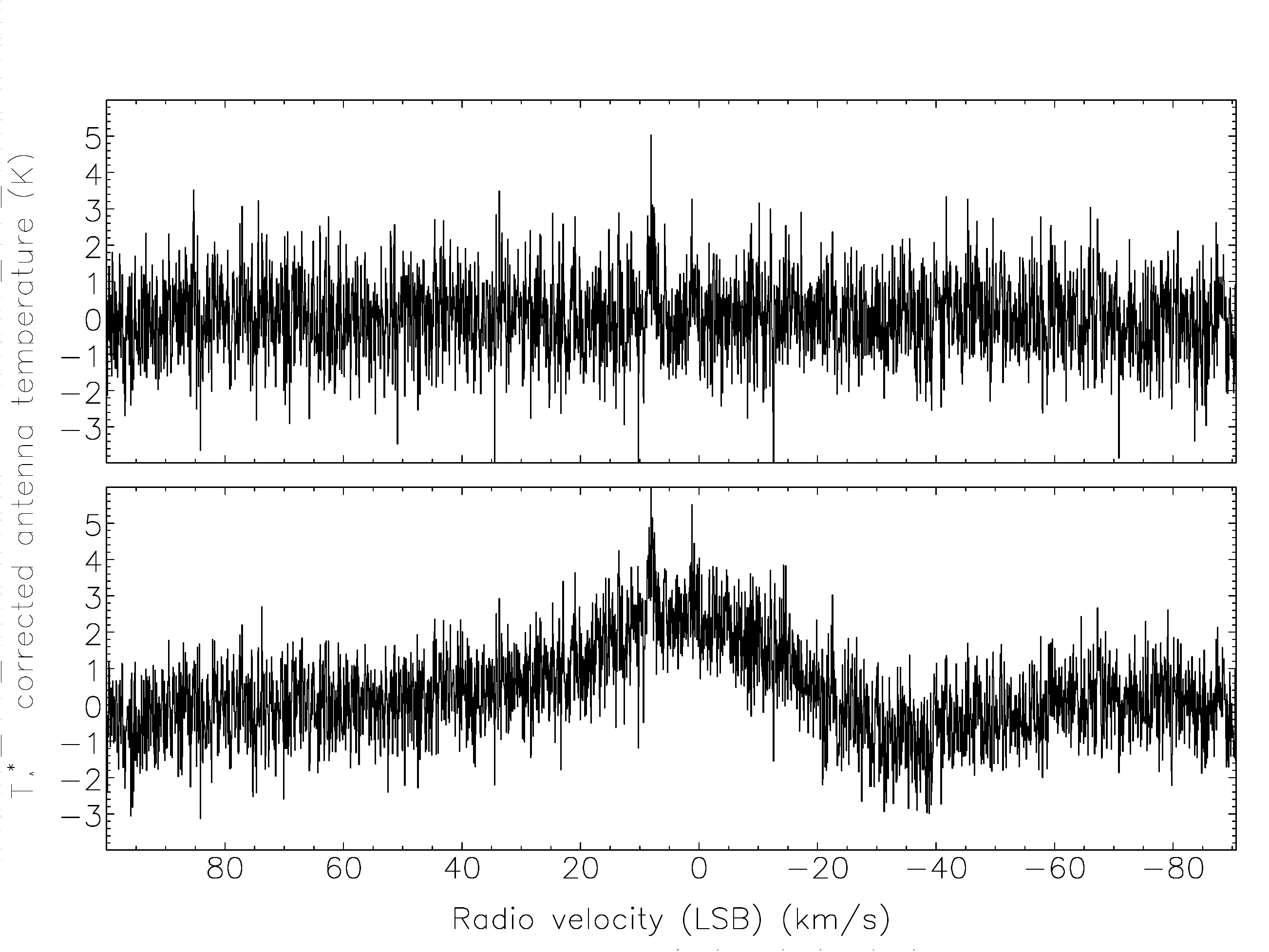}
\caption{Example of a reduced spectrum affected by a bad baseline.
  The lower panel shows a broad wave pattern, where a linear baseline
  subtraction would grossly overestimate the emission's flux.  The upper
  panel shows the corresponding spectrum after the application of
  non-linear baseline filtering described in \S\,\ref{sec:non-linearity}.}
\label{fig:badbase:spectrum}
\end{figure}

The spectra delivered by ACSIS/HARP can often include non-astronomical
signal arising from many sources, both local and external to the  JCMT. The
extraneous signal can appear in all detectors or just one, for a short
duration, or throughout an observation (see \S\ref{sec:obsmodes}).
This gives rise to artefacts in the spectral cube made by \makecube.  The anomalies
usually manifest as stripes or additional noise in the reduced
spectral cube.  Their presence at best degrades and sometimes dwarfs
the astronomical content.  The latter occurring with greater frequency
for early HARP observations.  In addition uneven baselines can lead
to poorly determined line fluxes (see Fig.~\ref{fig:badbase:spectrum}).

While many of the artefacts are readily visible in the raw time
series, and thus could be excised manually, some are subtle and easily
overlooked.  After a few years of operation of HARP, astronomers were
suspicious of all the data from a detector afflicted by anomalous signals,
choosing to exclude that detector's spectra completely, and thus
discarded perfectly good spectra in the presence of merely transient
interference.  The automated pipeline sought to address this
in a systematic fashion and to retain unaffected spectra.
This approach leads to more-uniform products within a survey.  Further
filters can be added as newly identified forms of bad spectra become
known.

\begin{figure}
\includegraphics[width=\columnwidth]{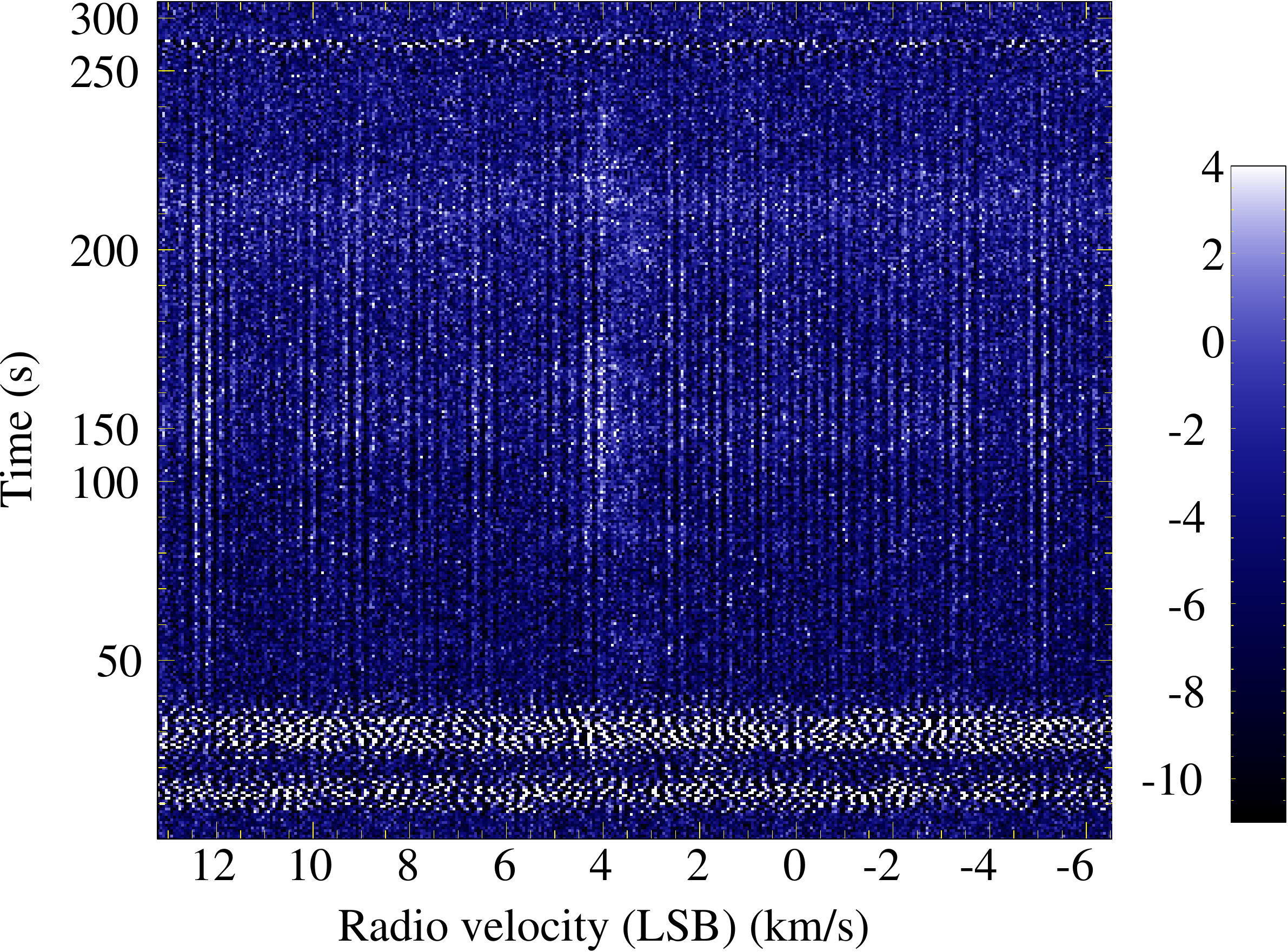}
\caption{Examples of the three main types of high-frequency
  interference depicted in spectral-time axes.  To the foot of this extract
  are bands of uncorrelated noise, and near the top is an isolated
  noisy spectrum.  Between lie spectra affected by ringing.
  The non-linear temporal axis arises because of intervals during
  which integrations of an off-source reference position are interspersed.}
\label{fig:badbase:highfreq}
\end{figure}

The bad baselines can be divided roughly into two classes:
high-frequency and low-frequency patterns.  The high-frequency
patterns are usually characterised by large-amplitude noise
arranged mostly in single isolated spectra or in bands comprising
around ten spectra.  Less frequently the noise manifests as
spiky spectra.  For the last two types the noise pattern phase shifts
between adjacent spectra.  In the first type there is beating in the
amplitudes.  A further form comprises weaker-amplitude striations
persistent over tens to 200 spectra, and it usually appears in
addition to the short-duration intense noise.  This correlated
`ringing' exhibits a bell-shape variation in intensity over time.
Fig.~\ref{fig:badbase:highfreq} presents the most-common forms.

\begin{figure*}
\includegraphics[width=\textwidth]{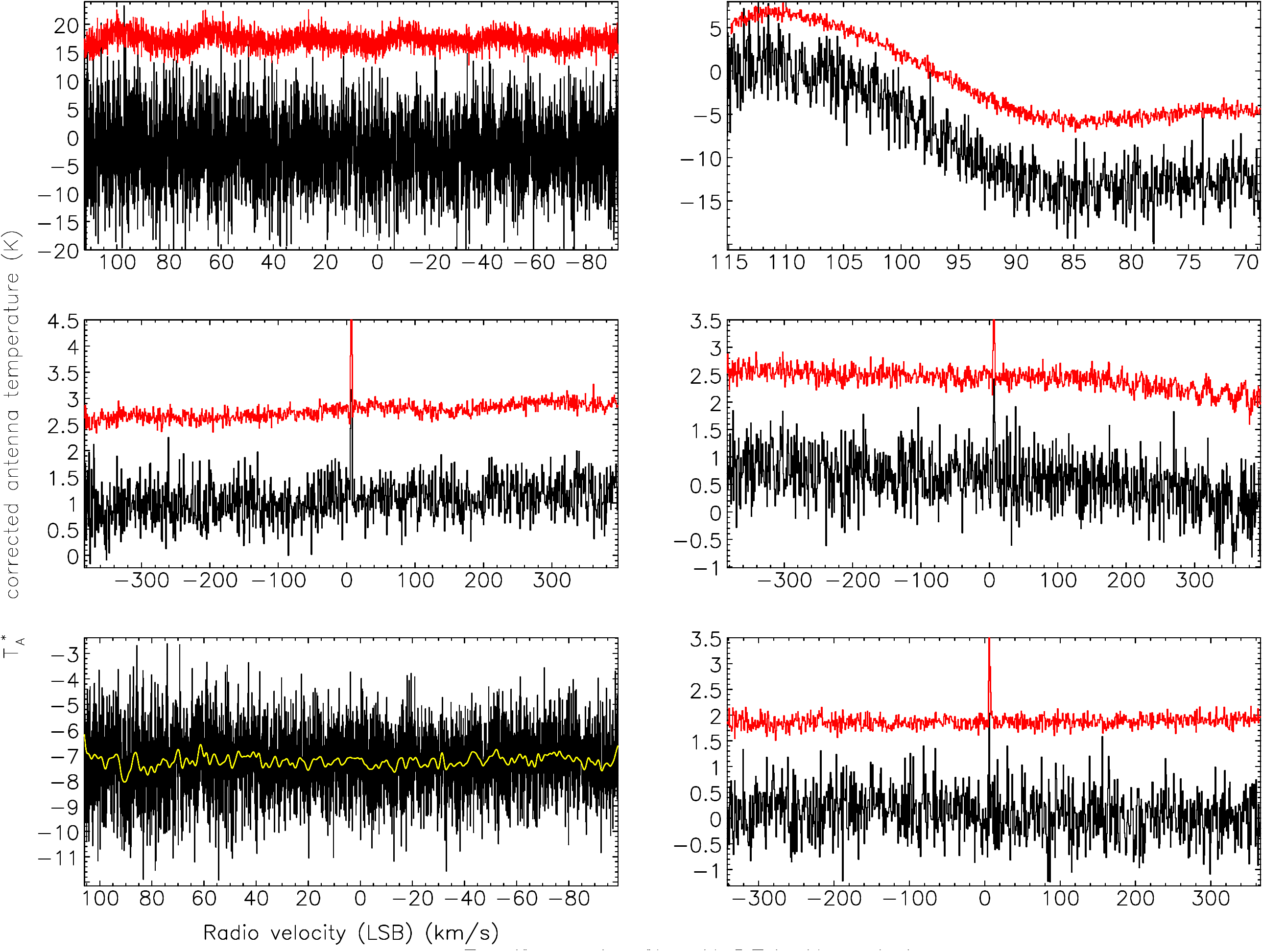}
\caption{Examples of low-frequency noise.  In each panel the
  black curve shows a single spectrum exhibiting a non-linear baseline
  of increasing degree from the lower to the top graphics.  A red curve
  is the average spectrum during an interval of bad behaviour, offset
  for clarity.  The yellow curve in the lower-left plot is a
  1.5\,km~s$^{-1}$ Gaussian smooth of the noisy spectrum showing a
  weak non-linearity towards the ends of the spectrum.}
\label{fig:badbase:interference}
\end{figure*}

\begin{figure}
\includegraphics[width=\columnwidth]{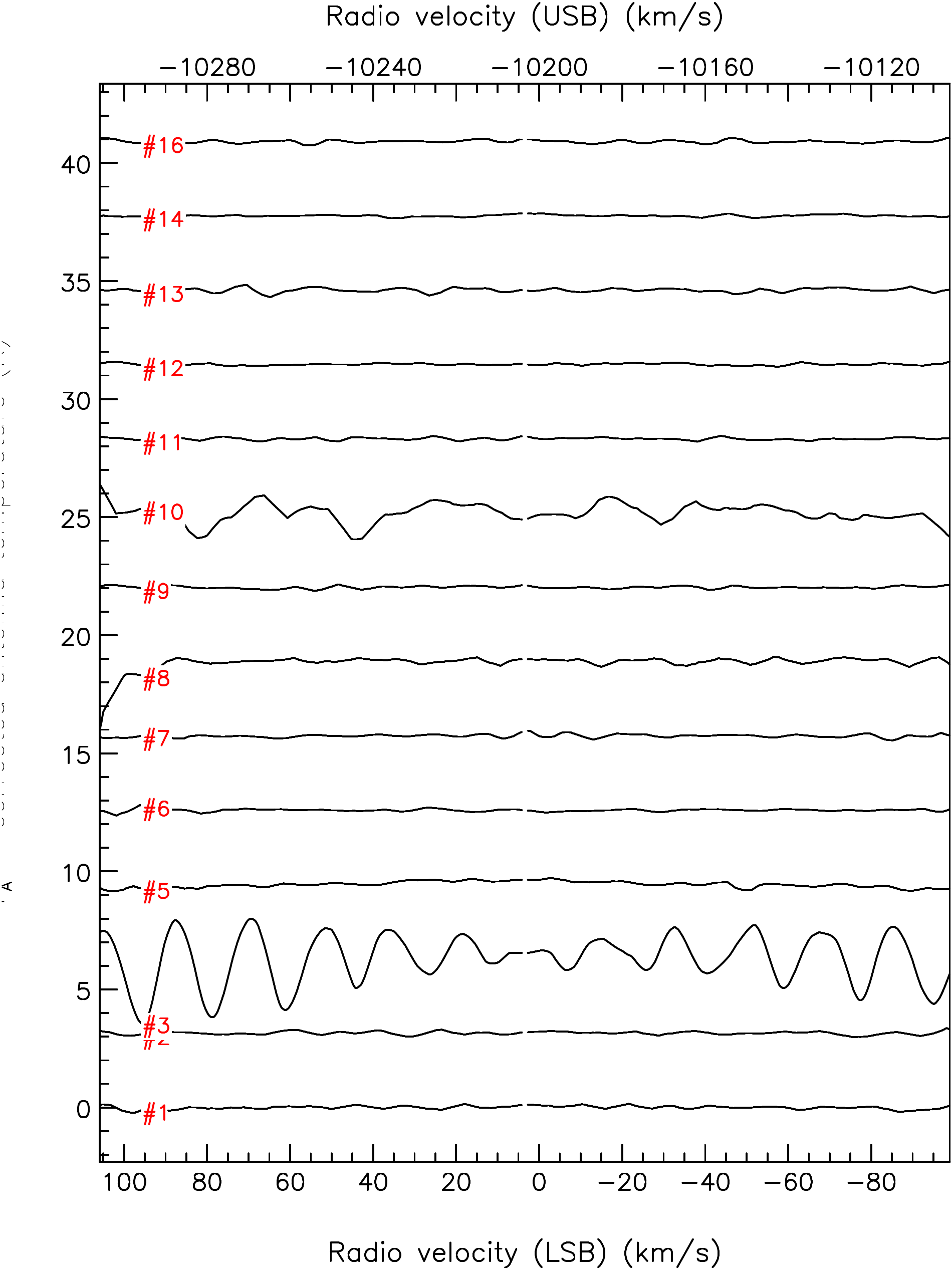}
\caption{Examples of low-frequency noise by detector.
  It shows time-averaged (clipped mean) spectra for each detector in
  which the third (labelled \#3) and ninth (labelled \#10) detector from the bottom
  exhibit strong global non-linear baselines.  Some other detectors
  have weaker non-linear baselines.  The central emission line
  is masked out and the noisy peripheries are excluded. The Y axis is
  the corrected  antenna temperature, in kelvin.}
\label{fig:badbase:interference_receptors}
\end{figure}

The low-frequency ripples tend to occur in time-series blocks that are
often visible because of baseline drift, but can apply to all spectra
for a detector.  They have a wide range of morphologies such as
sinusoids; irregular ripples; curved; and apparent emission initially
concentrated at two frequencies, but which disperse linearly in
frequency and fade with time reminiscent of fanned car headlight
beams seen from above.
Fig.~\ref{fig:badbase:interference} displays some examples.
Fig.~\ref{fig:badbase:interference_receptors} presents an example
observation where all the spectra are markedly non-linear for two
detectors.

The pipeline applies three steps in the quality-assurance stage:
\begin{itemize}
\item Laplacian filtering of high-frequency noise;
\item non-linearity detection for individual spectra; and
\item global non-linearity to reject whole detectors.
\end{itemize}

These are discussed in the following sections.

\subsubsection{Masking of High-Frequency Noise}

\begin{figure*}
\includegraphics[width=\textwidth]{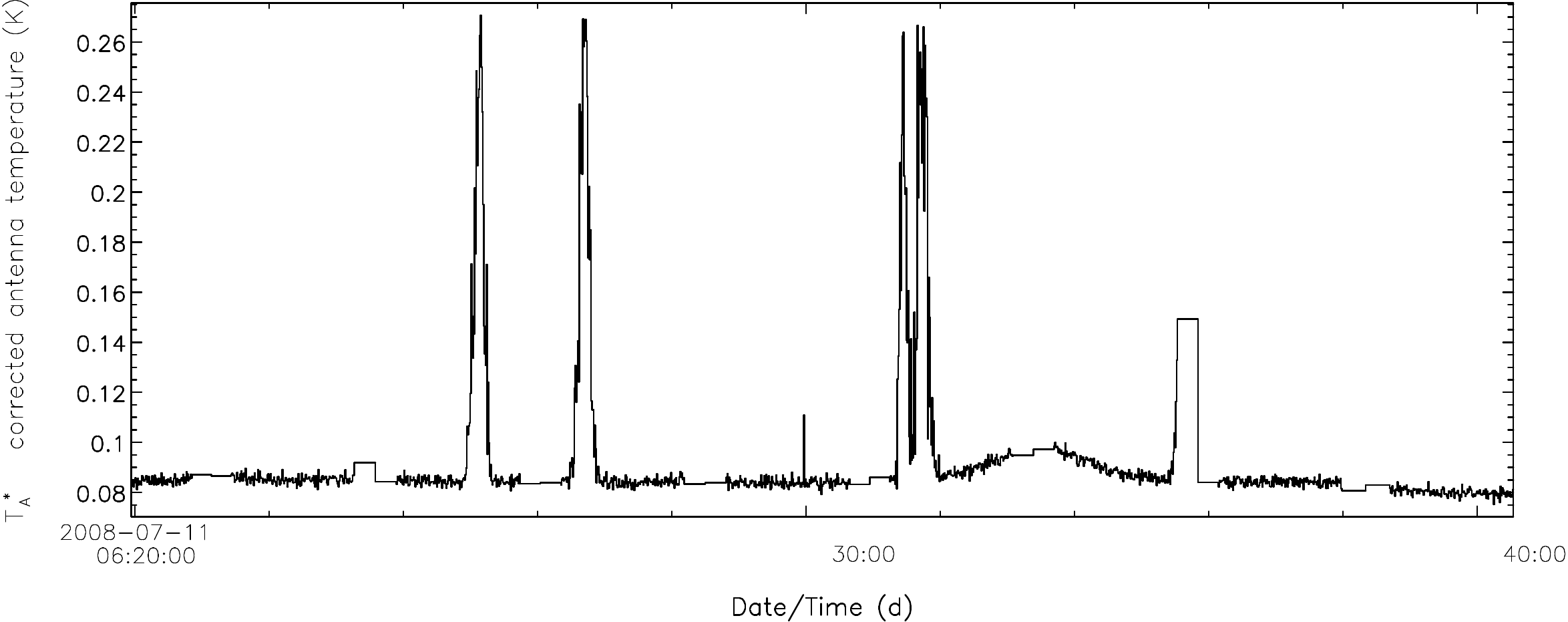}
\caption{An example raw edginess profile.  There are five broad
  peaks and one single-spectrum interference around 06:30.  Between
  06:32 and 06:35 is the much weaker signature of ringing.}
\label{fig:badbase:raw_edginess_profile}
\end{figure*}

The recipe applies a one-dimensional Laplacian edge filter to all the
spectra for each detector, after trimming the outer 15\,per~cent where noise
is always present.  This approximates to a difference-of-Gaussian
filter.  It next averages the rms `edginess' along the spectral axis
to form a profile through the time series.  An example profile is
shown in Fig.~\ref{fig:badbase:raw_edginess_profile}.  \cupid\
\findback\ subtracts any drifting background level ignoring the narrow
interference spikes.  Steps in the profile baseline are removed, where
possible.\footnote{Step correction currently invokes the \smurf\ \fixsteps\
application which was designed for long time series from the SCUBA-2
instrument \citep{2013MNRAS.430.2545C}, and in practice
requires hundreds of profile elements that are not always available.}
The final stage is to eject spectra whose rms edginess exceeds the
median level by a nominated number of clipped standard deviations.
Affected spectra are easily delineated.  However, the clipping can
leave residual spikes in the profile from the ramp up and down of the
interference signal.  This occurs when the standard deviation includes
both the noise and significant actual low-level variations, such as
caused by ringing, and hence raises the threshold too high.  The
algorithm applies a one- or two-element dilation to the excised
regions.  While this may throw away the odd good spectrum, it is more
than compensated by fewer artefacts in the reduced cube.

\begin{figure*}
\includegraphics[width=\textwidth]{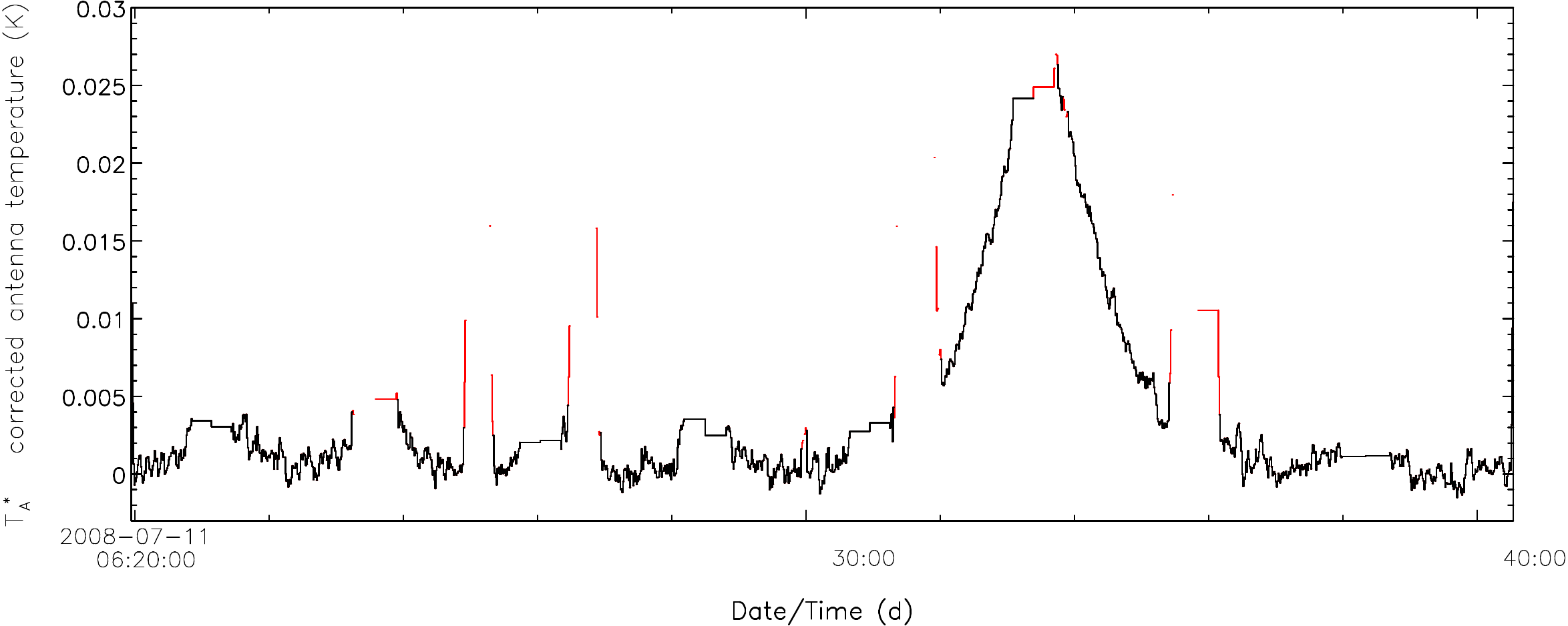}
\caption{The edginess profile after masking the short-duration
  interference, leaving the ringing signal.  The red is residual
  short-duration interference removed by the dilation.}
\label{fig:badbase:ringing_edginess_profile}
\end{figure*}

An optional second iteration removes most of the striation noise once
the pronounced edginess peaks are masked.  Determining the extent of
this ringing is problematic because of the noise, until we apply the
knowledge that the effect is correlated, which permits smoothing
along the time axis.
Fig.~\ref{fig:badbase:ringing_edginess_profile} shows that a ringing
signal can persist for longer than a mere visual inspection would
suggest, and there can also be weaker and shorter or periodic ringing
noise.  Using an estimate of the background and noise, the pipeline
then calls \cupid\ \findclumps\ to determine the location and extent
of the ringing.

At present the spectra affected ringing are masked.  Another approach
is to determine a normalised form of the ringing pattern and subtract
it for all affected spectra after using the edginess profile to scale
the intensity.  Thus more spectra would be combined into the reduced
cube.  Given the apparent periodicities, filtering in frequency
space might also prove effective.

\subsubsection{Non-linearity Filtering}
\label{sec:non-linearity}

The low-frequency ripple and wobbly baselines are addressed by
determining the non-linearity of each spectrum.  Since the baseline
is expected to be well fit by a straight line, the technique
measures the broad deviations of the smoothed baseline from a
straight-line fit.

First the recipe excludes non-baseline features that would dilute the
non-linearity signal.  These comprise a threshold to remove spikes and
mask the astronomical signal.  To exclude the astronomical emission
the recipe masks either in user-specified velocity ranges, or
determines the location of the emission unaided (see
\S\,\ref{sec:emission_detection}).

The recipe estimates the background level, effectively smoothing to
remove structure smaller than a nominated scale.  Next it fits linear
baselines to these and calculates the rms residuals to provide a
rectified signal.  Then it averages the signal along the spectral axis
to form a non-linearity profile through the time series for each good
detector.

The non-linear profiles are much noisier than the summed Laplacians
for the high-frequency interference, and discrimination is harder.  To
identify anomalous spectra the recipe reduces the noise to obtain a
smooth profile, correct for drifts or steps in the profile.  It
rejects spectra whose mean non-linearity exceeds the mean level above
a nominated number of clipped standard deviations.  The derived
standard deviation allows for positive skewness.  The final stage
is apply the mask of rejected spectra to the input cube.

The global non-linearity test is applied last so that a block of
transient highly deviant spectra will not cause the whole detector to
be rejected.  It operates in a similar fashion to the above.  It
diverges by determining a mean rms residual from non-linearity per
detector, from which it evaluates the median and standard deviation of
the distribution of mean rms residuals from the entire observation,
and performs iterative sigma clipping above the median to reject those
detectors whose deviations from linearity are anomalous.  There is a
tunable minimum threshold.

\subsubsection{Emission detection for non-linearity}
\label{sec:emission_detection}

When processing nightly observations for the JCMT Science Archive, the
location and extent of astronomical emission is usually unknown, yet
non-linear baselines need to be removed to generate reduced products
of acceptable fidelity.  The recipe uses an unsophisticated, but
effective scheme, to remove sufficient power from emission lines for
the non-linearity tests.

First it forms an integrated spectrum for the observation by averaging
in time, and then forming the clipped mean along detectors to exclude
the effects of strong non-linearity in one or two detectors.  The
representative spectrum has a linear baseline subtracted in the
automatic mode of \mfittrend (see \S\,\ref{sec:mfittrend}) to remove
any pronounced slope.  This spectrum is smoothed with a 51-element
Gaussian kernel to narrow the histogram peak of baseline values, and
make emission more prominent.  Then follows a multi-scale iterative
approach using histograms and \mfittrend\ to progressively improve the
baseline subtraction by excluding the outliers.  It starts with a
smoothing box one eighth the width of the spectrum and halving the box
at each iteration.  For each iteration the recipe forms a truncated
histogram whose mode and standard deviation are estimated, the latter
allowing for the positive skew from the astronomical signal.  Then the
recipe masks positive outliers (defaults to a 4-sigma clip).  Either
\findback\ or a block smooth using the current box size is subtracted
to give a flattish baseline for \mfittrend\ in automatic mode to
determine the baseline regions and hence the remaining emission.
\mfittrend\ on its own can have difficulty separating broad-line
emission from baseline.  One iteration is usually sufficient.

\subsubsection{Results}

\begin{figure*}
\includegraphics[width=\textwidth]{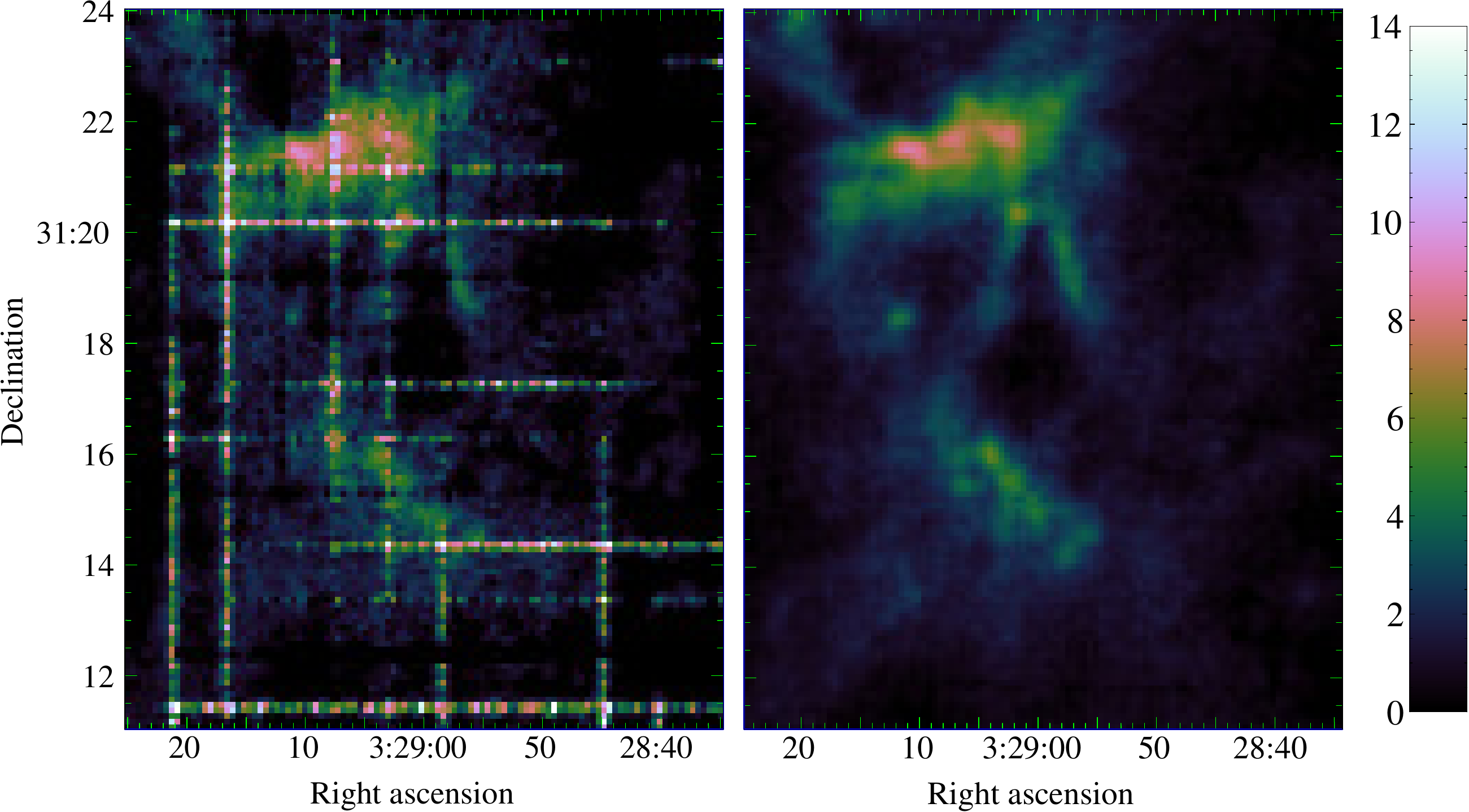}
\caption{A bad-spectra comparison displayed in an integrated intensity map.
  The left panel shows a map without bad-baseline removal.  The right
  panel shows the same data processed using the high- and
  low-frequency noise filters.  While the peak astronomical signal in
  the image is approximately 8, the presence of artefacts cause nearly
  two percent of all pixels to appear brighter, reaching a maximum over 17.
  No flatfield correction has been applied. These data were from
  project M06BGT02; observations 65 to 67 on 2007 July 28 and
  observations 8, 9, 11, 12, 15, 16, 22 and 23 from 2007 December 17.}
\label{fig:badbase:results}
\end{figure*}

The methods appear highly effective at cleaning the pipeline products,
as can be seen in Fig.~\ref{fig:badbase:results}. \citep[See][for
details of earlier reductions of these data.]{2010MNRAS.401..455C} The
filtering has been used to re-reduce two surveys and many other data
sets. This includes at least one that had originally failed quality
assurance, but now has been used for science
\citep{2013ApJ...767..126S}.

\subsection{Flatfielding}
\label{sec:flat}

HARP data from its early years seemed to have problems with the
relative calibration of the detectors.  A self flatfielding algorithm
was developed that relied on the detectors on average seeing the same
signal for large scan maps of molecular clouds
\citep{2010MNRAS.401..455C}, and this had some success in removing
striping from integrated intensity images.  For an observation the
Curtis technique creates spectral cubes for each detector
independently and compares the integrated fluxes over the emission
line.  This approach is, however, only successful where the signal
being compared extends across a significant portion of the spatial
plane to mitigate against different detectors observing different
flux.  Self flatfielding is also limited to data whose signal-to-noise
ratio of the total flux in each detector permits relative
sensitivities to better than about 5 percent.  Since these constraints
are often not true, the pipeline recipes by default do not apply
flatfield corrections.

When a user does request that a flatfield be applied, the recipe
segregates all the observations supplied to the recipe by date or
individually.  Although the detector-to-detector response has been
known to vary during the course of a night, far more often it is
stable.  Thus combining observations, such as the two directions of a
weave, or any repeat observations through a night, permits a better
determination of the flat field.  Such a flatfield can also be applied
to low-signal data taken on the same night too.  The flux is summed
either over a parameter-controlled spectral range, or from the
group-determined emission map.  Then the fluxes are normalised by the
flux of the reference detector, or its reserve, should the reference
detector be disabled or has failed quality assurance.

\begin{figure*}
\includegraphics[width=\textwidth]{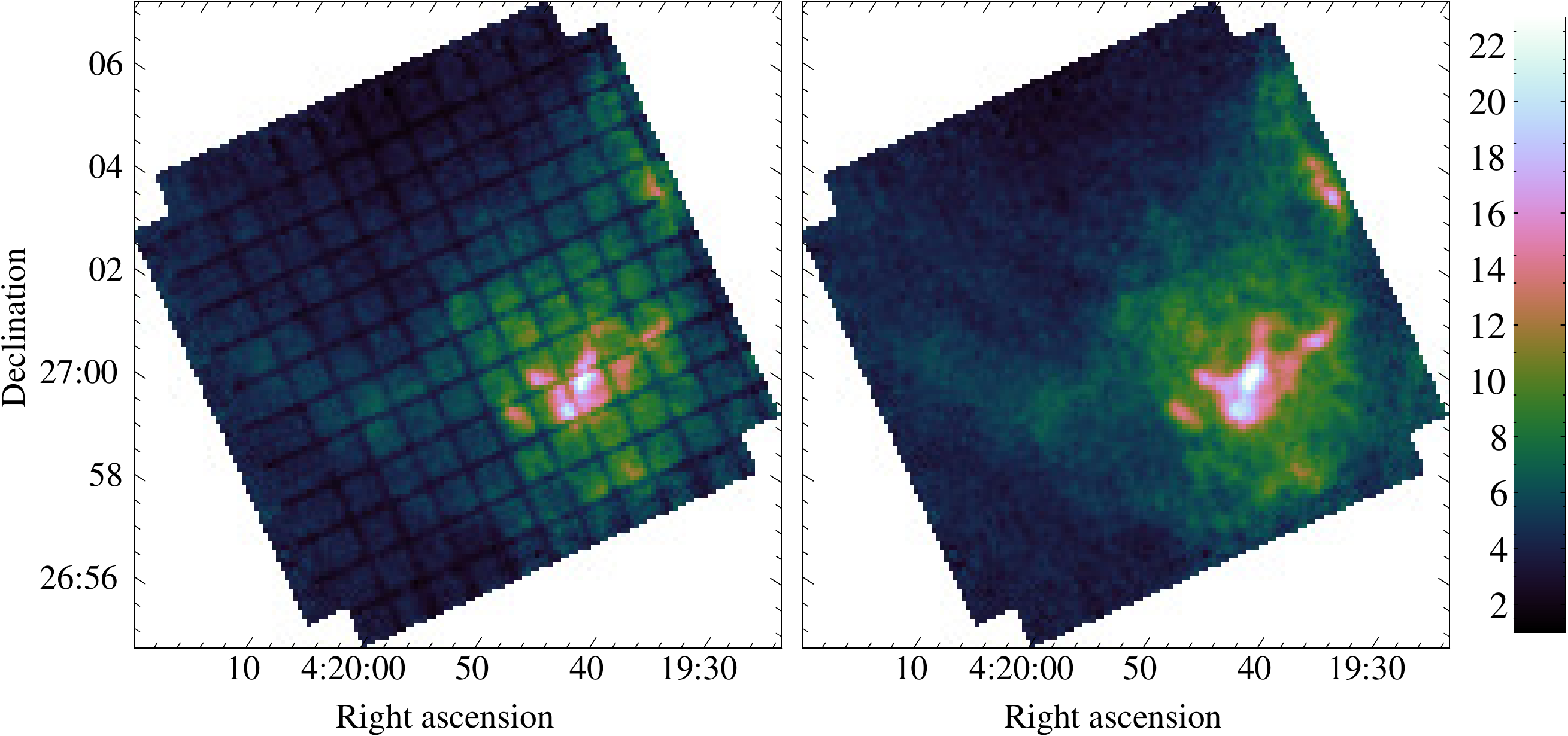}
\caption{A flat-field comparison displayed in an integrated intensity
  map.  The left panel shows a integrated map without correction for
  detector-to-detector performance.  The right panel shows the same
  data processed but also applying the sum method to flat-field.  Both
  images are scaled to the same intensity limits.  Most occurrences
  exhibit weaker differences in detector responsivity than present in
  this example. These data were from observations 27 and 28 from 2008
  November 12, project  MJLSG17.}
\label{fig:flatfield:results}
\end{figure*}

With appropriate data this works well (see
Fig.~\ref{fig:flatfield:results} for an example).  However, weak
residual graticule patterns can remain if the reference field contains
a compact source whose line emission falls within the same spectral
limits as the flux summation and is not observed uniformly by all
detectors.  Fourier techniques can be used to filter such patterns and
improve the cosmetic appearance of cubes \citep{2015MNRAS.447.1996W}.

We found one occurrence in early data where the flat field broke down
because there appeared to be non-linearity in the signal.  Different
flatfield ratios were needed for the bright regions compared with near
the sky level.  The pipeline makes no provision for such data.

We explored other methods to mitigate against the limitations with
little or inconsistent success.  For instance, using the peaks of the
histogram of the ratios of pixel by pixel was biased by the noise,
even if comparisons were restricted to the spectral range of the
astronomical emission.  None proved better than the Curtis formula.

\subsection{Quality-Assurance Parameters \label{sec:qa}}

\begin{table}
  \caption{Summary of the quality-assurance parameters supported by
    the pipeline. More details can be found in \citep{2008JCMTLSQA}.}
\label{tab:qa:params}
\begin{tabular}{ll}
BADPIX\_MAP  & Percentage of bad spatial pixels\\
   & in output product\\
CALINTTOL & Percentage discrepancy allowed\\
    &  in calibrator integrated intensity \\
CALPEAKTOL & Percentage discrepancy allowed\\
    & in calibrator peak\\
FLAGTSYSBAD & Percentage of data allowed \\
   & to be flagged due to T$_{sys}$\\
GOODRECEP & Number of functioning detectors\\
RESTOL & Tolerance on residuals of baseline \\
   & region after baseline subtraction\\
RESTOL\_SM & Variation of baseline residuals \\
  & over restricted range\\
RMSTOL & Consistency check comparing\\
  & T$_{sys}$ with spectrum rms\\
RMSVAR\_MAP & Percentage variation of\\
  & rms noise across map \\
RMSVAR\_RCP &  Percentage average rms detectors\\
  & are allowed to vary from each other\\
RMSVAR\_SPEC & Percentage variation of RMS\\
  & across spectrum \\
TSYSBAD &  Maximum allowable T$_{sys}$\\
TSYSMAX & Threshold for  average T$_{sys}$ \\
  & allowed for a detector\\
TSYSVAR &  Maximum allowed variation of\\
  & T$_{sys}$ for a single detector\\
\end{tabular}
\end{table}

The survey teams required a comprehensive set of quality-assurance
testing to ensure that the data quality is consistent for the duration
of the surveys \citep{2008JCMTLSQA}. A number of QA tests were added
to the pipeline and a full list is shown in Table
\ref{tab:qa:params}. Some of these QA parameters are designed to be
run on spectral line standards observations prior to starting science
observations to determine whether the system is configured properly
and to allow the observer to decide which, if any, of the projects can
be observed next. There are also QA tests designed to look at the
time-series data and others that analyse the map/cube products.

\subsection{Alternative Recipes}
\label{sec:alt}

It is not possible or even desirable for a single data reduction
technique to apply to many different types of sources and science
goals. For that reason a number of different recipes are made
available and these can be chosen by the observer in the Observing
Tool, overridden later on the command-line or by using a
configuration file at the JSA.

\subsubsection{Gradient}

This is the standard recipe optimized for nearby galaxies or other
objects where there can be a velocity gradient across
the field of view. This velocity gradient is a major motivation for the
automated detection of baseline regions as it allows you to maximize
the baseline region rather than supplying a simple range that
encompasses all the data. The major difference with the narrow-band
recipe below is that the smoothing is biased towards spectral smoothing
rather than spatial smoothing.

\subsubsection{Narrow line}

This recipe is used for observations of objects with relatively narrow
lines and a small velocity gradient (compared with the total observed
band). This applies to many Galactic targets. Smoothing is biased
towards spatial smoothing.

\subsubsection{Broad line}

Active galaxies often have broad lines of several hundreds of km/s so
this recipe is tuned to be less aggressive for automatic baseline
subtraction than the standard recipe that is designed for nearby galaxies.
An early form of this recipe, in the form of a standalone script, was used
for the initial Nearby Galaxy Survey data release and is documented in
\citet{2010ApJ...714..571W}.

\section{Processing of Historical JCMT Data}

\begin{figure*}
\centering
\includegraphics[width=\textwidth]{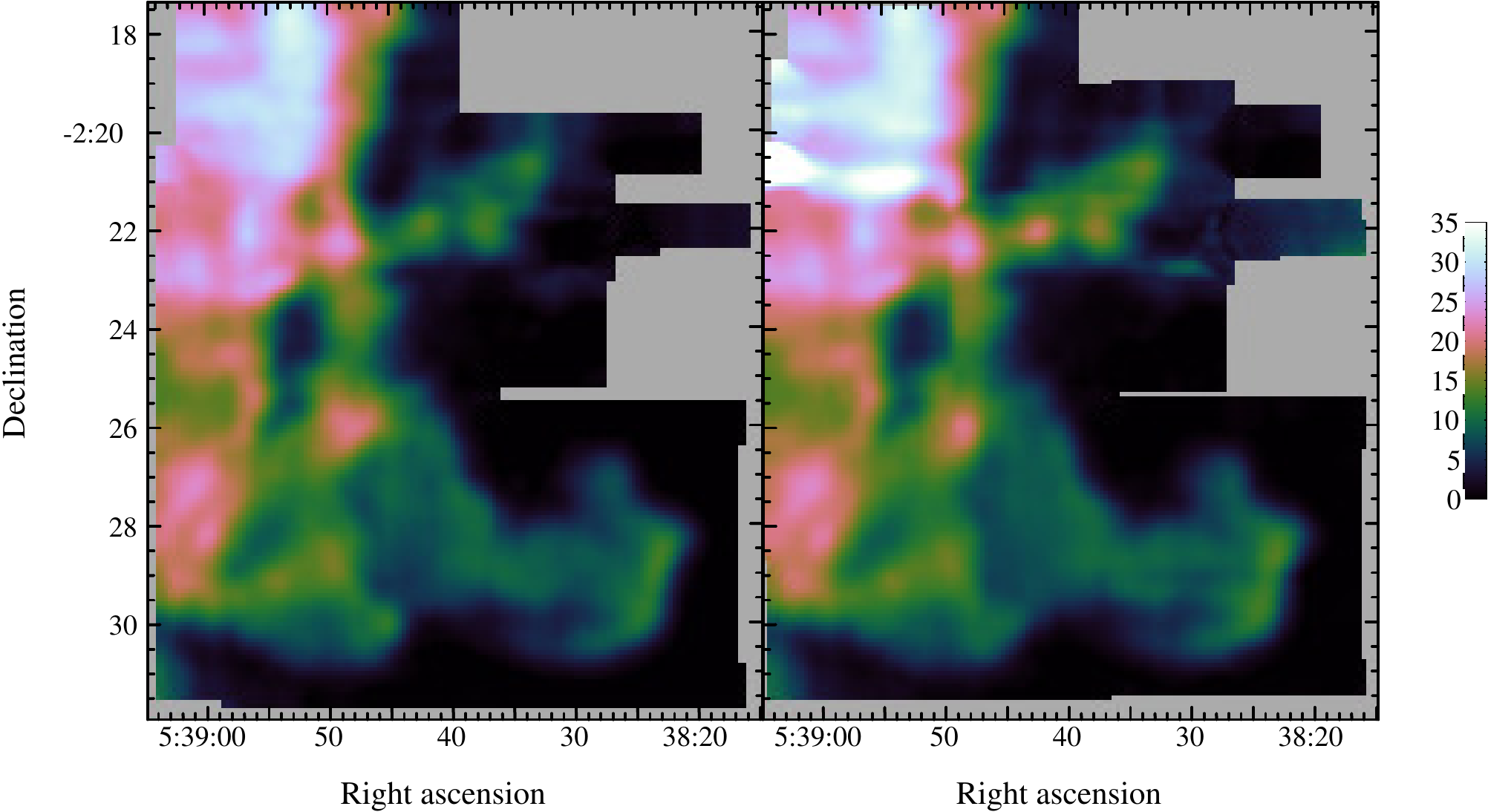}
\caption{Left is an integrated intensity image created from
  a cube made interactively using \specx. Right is an
  integrated intensity image made from a pipeline reduction of the
  same raw data. Intensity scales are 0 to 35 K\,km\,s$^{-1}$.}
\label{fig:hhcmp}
\end{figure*}

Until ACSIS was delivered in 2005, heterodyne data were taken with a
variety of backend systems including the Acousto-Optical Spectrometer
(AOSC) and the Dutch Autocorrelation Spectrometer (DAS)
\citep{1986SPIE..598..134B}. These backends wrote data in the
Global Section Data (GSD) format \citep[e.g.,][]{GSD1999,GSD2015} which was
understood by the \specx\ data reduction package. To ensure that as many
of these historical data as possible are made available to the community
in a usable format, we have developed an extension to the \smurf\
package called \gsdacsis, which converts the legacy data to the
newer ACSIS data format. This enables the legacy data to be re-processed using
the modern data reduction pipeline and automatically leads to these
products being easily available from the JSA.

The main difficulty in supporting legacy data in the pipeline related
to the DAS splitting the bandwidth into many overlapping
sub-bands. ACSIS data only ever had included two sub-bands for hybrid-mode
observations and so the sub-band merging algorithm had to be
extended to support arbitrary numbers of sub-bands.  In addition the
historical data are dominated by single-detector instruments and most
observations generated relatively few spectra, and few repeats of the
same area of sky (depth in the first pass was preferred over short
integration times but many repeats). This sometimes
restricts the benefits that can be obtained by using a pipeline
designed for focal plane arrays.

We tested the pipeline on one of the largest data sets from the
historical era. Observations of the Horsehead nebula were an
observatory backup project from 1995 to 1997. The $^{13}$CO
$J=2\rightarrow 1$ component of the project consisted of approximately
14,000 spectra from 154 observations spread over 12 nights using the
single-detector receiver RxA2 \citep{1992IJIMW..13..647D}. Reducing
the data with \specx\ was an involved process and preliminary results
were presented in \citet{2001AAS...19915601S}. For this test the GSD data were
downloaded from the CADC and converted to ACSIS format.

For these observations there were two major difficulties. The data
were taken in raster scan mode with over-sampling in the scan
direction (to prevent beam-smearing) and Nyquist sampling between
rows. For the JCMT beam at this frequency this corresponds for
5\,arcsec and 10\,arcsec spacing. The pipeline has not been optimized
for this observing scheme although it correctly selected a 5\,arcsec
pixel size. This meant that half of the map contained bad pixels and
so an additional interpolation routine was required before the final
integrated intensity image could be calculated. A consequence of the
large number of flagged pixels in the map was that the quality
assurance test associated with bad-pixel map fraction had to be
relaxed significantly.

For some of the observations a bad reference position had been chosen,
which leads to absorption features in some spectra. These data were
used in the manual reduction because a long observation was taken on
this reference position along with a new ``off'' position and the
spectra corrected. Doing this in an automated fashion is difficult
without more investigation into the related observations as GSD data
did not record the location of the reference position in the
header. For the purposes of the test these observations were not used.

The results can be seen in Fig.\ \ref{fig:hhcmp}. The \specx\ ``manual''
reduction does look better than the automated reduction, especially in
the region in the north east where some contamination still seems to
be present. It seems feasible to assume that some of this can be
improved upon by writing recipes specifically targeted at legacy data
but the proof of concept is encouraging.

\section{Conclusion}

With many thousands of spectra from a single observation it is
impractical to examine every spectrum manually. The data reduction
scheme described here is used continually at the JSA
\citep{2011ASPC..442..203E,2014SPIE9152-93} for daily and project processing and the
tuned recipes are now generating the primary products from the
heterodyne part of the Gould's Belt JCMT Legacy Survey
\citep{2007PASP..119..855W} and the CO High Resolution Survey
(COHRS), also from the JCMT \citep{2013ApJS..209....8D}.

The algorithms and approach described in this paper should be
applicable to other heterodyne instrumentation and are not
JCMT-specific. The implementation using ORAC-DR can only be used with
raw data written using the JCMT data model
(Appendix~\ref{sec:rawdata}). Some work has been done on software to
import data from Supercam \citep{2012SPIE.8452E..04K} and NANTEN2
SMART \citep{2008stt..conf..488G} into the JCMT format and tests are
ongoing. We hope that this work will influence the data acquisition
and data processing plans for other observatories building large focal
plane arrays. Accurate metadata is a critical component when
attempting to automate the reduction of thousands of spectra and it is
no longer acceptable for the data reduction software to have to guess
important information. For example, some telescopes report the
position at the start of the row but not the position during a scan,
and assume that the position of each spectra can be derived by looking
at the integration time. Trusting that the telescope behaved according
to the guesses of the software can work when the telescope is moving
slowly and someone is examining each spectrum but this approach is not
scalable.

The iterative pipeline processing described in this paper demonstrates
the possibilities for advanced heterodyne cube reconstruction if we
begin to use techniques more akin to those used by iterative
map-makers for bolometer cameras
\citep[e.g.,][]{2013MNRAS.430.2545C}. The next step is to explicitly
embrace such techniques, building up explicit models of the
astronomical emission and baselines and enhance this to have models
involving knowledge of which detectors have related local oscillators
and which detectors come from the same backend hardware. This latter
facility may be important as more and more detectors are added to
focal plane arrays and would be similar to dealing with readout issues
of bolometer arrays. Such an iterative cube-maker, coupled with novel
scanning strategies, may lead to a fundamental modification of
heterodyne observing modes where slow instrumental drifts can be
tracked without having to visit a reference sky position regularly
through an observation. It may be sufficient to measure the reference
position at the start and end of the observation and then model the
drifts during data reduction. This would result in significantly more
efficient observing modes, similar to that obtained when continuum
instruments moved from a chopping secondary to a total power
configuration.

\section*{Acknowledgments}

The James Clerk Maxwell Telescope has historically been operated by
the Joint Astronomy Centre on behalf of the Science and Technology
Facilities Council of the United Kingdom, the National Research
Council of Canada and the Netherlands Organisation for Scientific
Research. This work was funded by the Science and Technology Facilities
Council. We thank the many JCMT support scientists and survey
scientists who have tested the pipeline. In particular we thank
Jessica Dempsey, Holly Thomas, Jan Wouterloot, Jane Buckle and
Jennifer Hatchell. We thank Doug Johnstone for useful comments on a
draft of this paper. We also thank Jennifer Balfour and
Vincent Tilanus for their work on \gsdacsis\ and
G\"{o}ran Sandell for providing us with the \specx\ reductions of the Horsehead
Nebula. This research used the facilities of the Canadian Astronomy
Data Centre operated by the National Research Council of Canada with
the support of the Canadian Space Agency. This research has made use
of NASA's Astrophysics Data System.

This work was built on the Starlink Software Collection, which was
developed by the Starlink Project until 2005
\citep{1982QJRAS..23..485D,2005ASPC..347...22D,2008ASPC..394..650C}
and then opened up to the community. The source code for the Starlink
software (\ascl{1110.012}) and ORAC-DR is open-source and is
available on GitHub\footnote{\url{https://github.com/Starlink}}.

\appendix

\section{JCMT Heterodyne Raw Data Model}
\label{sec:rawdata}

Raw data files at JCMT are written to HDS format files
\citep[e.g.,][\ascl{1502.009}]{2015HDS} using the extensible \emph{N}-Dimensional
Data Format \citep[NDF;][\ascl{1411.023}]{2015NDF}. The spectral data are stored in a
three-dimensional data array dimensioned by spectrum channels/frequency, detector
number and time. In addition to a FITS-style header, three extensions
are used to describe the data. The \texttt{JCMTOCS} extension contains
a full description of the requested observation in XML format. The
\texttt{ACSIS} extension describes the individual detectors including
their positions in the focal plane, their names, and the system and
receiver temperatures. The \texttt{JCMTSTATE} extensions contains
time-varying information associated with each time step in the primary
data array. This primarily includes the telescope tracking position
and acquisition time of each spectrum (using TAI), but also includes
the local oscillator settings; environmental parameters, such as
temperature and air pressure; and the position of the secondary mirror
(only needed for jiggle observing modes).  A complete description of
the heterodyne data file format, including a full list of header
parameters, can be found in \citet{OCS_ICD_022}.


\bsp	
\label{lastpage}

\begin{thebibliography}{}
\makeatletter
\relax
\def\mn@urlcharsother{\let\do\@makeother \do\$\do\&\do\#\do\^\do\_\do\%\do\~}
\def\mn@doi{\begingroup\mn@urlcharsother \@ifnextchar [ {\mn@doi@}
  {\mn@doi@[]}}
\def\mn@doi@[#1]#2{\def\@tempa{#1}\ifx\@tempa\@empty \href
  {http://dx.doi.org/#2} {doi:#2}\else \href {http://dx.doi.org/#2} {#1}\fi
  \endgroup}
\def\mn@eprint#1#2{\mn@eprint@#1:#2::\@nil}
\def\mn@eprint@arXiv#1{\href {http://arxiv.org/abs/#1} {{\tt arXiv:#1}}}
\def\mn@eprint@dblp#1{\href {http://dblp.uni-trier.de/rec/bibtex/#1.xml}
  {dblp:#1}}
\def\mn@eprint@#1:#2:#3:#4\@nil{\def\@tempa {#1}\def\@tempb {#2}\def\@tempc
  {#3}\ifx \@tempc \@empty \let \@tempc \@tempb \let \@tempb \@tempa \fi \ifx
  \@tempb \@empty \def\@tempb {arXiv}\fi \@ifundefined
  {mn@eprint@\@tempb}{\@tempb:\@tempc}{\expandafter \expandafter \csname
  mn@eprint@\@tempb\endcsname \expandafter{\@tempc}}}

\bibitem[\protect\citeauthoryear{Bell et~al.,}{Bell
  et~al.}{2014}]{2014SPIE9152-93}
Bell G.~S.,  et~al., 2014, in Chiozzi G.,  Radziwill N.~M.,  eds,  \procspie\
  Vol. 9152, {Software and Cyberinfrastructure for Astronomy III}. p. 91522J,
  \mn@doi{10.1117/12.2054983}

\bibitem[\protect\citeauthoryear{Berry}{Berry}{2015}]{2015FW}
Berry D.~S.,  2015, \mn@doi [Astronomy \& Computing]
  {10.1016/j.ascom.2014.11.004}, 10, 22

\bibitem[\protect\citeauthoryear{{Berry} \& {Jenness}}{{Berry} \&
  {Jenness}}{2012}]{2012ASPC..461..825B}
{Berry} D.~S.,  {Jenness} T.,  2012, in {Ballester} P.,  {Egret} D.,
  {Lorente} N.~P.~F.,  eds,  ASP Conf. Ser. Vol. 461, Astronomical Data
  Analysis Software and Systems XXI. p.~825 (\mn@eprint {arXiv} {1210.5483})

\bibitem[\protect\citeauthoryear{{Berry}, {Reinhold}, {Jenness}  \&
  {Economou}}{{Berry} et~al.}{2007}]{2007ASPC..376..425B}
{Berry} D.~S.,  {Reinhold} K.,  {Jenness} T.,   {Economou} F.,  2007, in {Shaw}
  R.~A.,  {Hill} F.,   {Bell} D.~J.,  eds,  ASP Conf. Ser. Vol. 376,
  Astronomical Data Analysis Software and Systems XVI. p.~425

\bibitem[\protect\citeauthoryear{{Bos}}{{Bos}}{1986}]{1986SPIE..598..134B}
{Bos} A.,  1986, in {Kollberg} E.,  ed.,  Proc.~SPIE Vol. 598, Instrumentation
  for submillimeter spectroscopy. pp 134--140, \mn@doi{10.1117/12.952332}

\bibitem[\protect\citeauthoryear{{Buckle} et~al.,}{{Buckle}
  et~al.}{2009}]{2009MNRAS.399.1026B}
{Buckle} J.~V.,  et~al., 2009, \mn@doi [MNRAS]
  {10.1111/j.1365-2966.2009.15347.x}, \href
  {http://adsabs.harvard.edu/abs/2009MNRAS.399.1026B} {399, 1026}

\bibitem[\protect\citeauthoryear{Chapin, Gibb, Jenness, Berry  \&
  Tilanus}{Chapin et~al.}{2013a}]{SUN258}
Chapin E.,  Gibb A.~G.,  Jenness T.,  Berry D.~S.,   Tilanus R.,  2013a,
  Starlink User Note~258, SMURF -- the Sub-Millimetre User Reduction Facility.
Joint Astronomy Centre

\bibitem[\protect\citeauthoryear{{Chapin}, {Berry}, {Gibb}, {Jenness}, {Scott},
  {Tilanus}, {Economou}  \& {Holland}}{{Chapin}
  et~al.}{2013b}]{2013MNRAS.430.2545C}
{Chapin} E.~L.,  {Berry} D.~S.,  {Gibb} A.~G.,  {Jenness} T.,  {Scott} D.,
  {Tilanus} R.~P.~J.,  {Economou} F.,   {Holland} W.~S.,  2013b, \mn@doi
  [MNRAS] {10.1093/mnras/stt052}, \href
  {http://adsabs.harvard.edu/abs/2013MNRAS.430.2545C} {430, 2545}

\bibitem[\protect\citeauthoryear{{Chrysostomou}}{{Chrysostomou}}{2010}]{2010HiA....15..797C}
{Chrysostomou} A.,  2010, \mn@doi [Highlights of Astronomy]
  {10.1017/S1743921310011750}, \href
  {http://adsabs.harvard.edu/abs/2010HiA....15..797C} {15, 797}

\bibitem[\protect\citeauthoryear{Cressie}{Cressie}{1990}]{1990Cressie}
Cressie N.,  1990, \mn@doi [Mathematical Geology] {10.1007/BF00889887}, 22, 239

\bibitem[\protect\citeauthoryear{{Cunningham}, {Hayward}, {Wade}, {Davies}  \&
  {Matheson}}{{Cunningham} et~al.}{1992}]{1992IJIMW..13.1827C}
{Cunningham} C.~T.,  {Hayward} R.~H.,  {Wade} J.~D.,  {Davies} S.~R.,
  {Matheson} D.~N.,  1992, \mn@doi [\ijimw] {10.1007/BF01011325}, \href
  {http://adsabs.harvard.edu/abs/1992IJIMW..13.1827C} {13, 1827}

\bibitem[\protect\citeauthoryear{Currie \& Berry}{Currie \&
  Berry}{2013}]{SUN95}
Currie M.~J.,  Berry D.~S.,  2013, Starlink User Note~95, KAPPA -- Kernel
  Application Package.
Joint Astronomy Centre

\bibitem[\protect\citeauthoryear{{Currie}, {Draper}, {Berry}, {Jenness},
  {Cavanagh}  \& {Economou}}{{Currie} et~al.}{2008}]{2008ASPC..394..650C}
{Currie} M.~J.,  {Draper} P.~W.,  {Berry} D.~S.,  {Jenness} T.,  {Cavanagh} B.,
    {Economou} F.,  2008, in {Argyle} R.~W.,  {Bunclark} P.~S.,   {Lewis}
  J.~R.,  eds,  ASP Conf. Ser. Vol. 394, Astronomical Data Analysis Software
  and Systems XVII. p.~650

\bibitem[\protect\citeauthoryear{{Currie}, {Berry}, {Jenness}, {Gibb}, {Bell}
  \& {Draper}}{{Currie} et~al.}{2014}]{2014ASPC..485..391C}
{Currie} M.~J.,  {Berry} D.~S.,  {Jenness} T.,  {Gibb} A.~G.,  {Bell} G.~S.,
  {Draper} P.~W.,  2014, in {Manset} N.,  {Forshay} P.,  eds,  \aspconf Vol.
  485, Astronomical Data Analysis Software and Systems XXIII. p.~391

\bibitem[\protect\citeauthoryear{{Curtis}, {Richer}  \& {Buckle}}{{Curtis}
  et~al.}{2010}]{2010MNRAS.401..455C}
{Curtis} E.~I.,  {Richer} J.~S.,   {Buckle} J.~V.,  2010, \mn@doi [MNRAS]
  {10.1111/j.1365-2966.2009.15658.x}, \href
  {http://adsabs.harvard.edu/abs/2010MNRAS.401..455C} {401, 455}

\bibitem[\protect\citeauthoryear{{Davies}, {Cunningham}, {Little}  \&
  {Matheson}}{{Davies} et~al.}{1992}]{1992IJIMW..13..647D}
{Davies} S.~R.,  {Cunningham} C.~T.,  {Little} L.~T.,   {Matheson} D.~N.,
  1992, \mn@doi [\ijimw] {10.1007/BF01010688}, \href
  {http://adsabs.harvard.edu/abs/1992IJIMW..13..647D} {13, 647}

\bibitem[\protect\citeauthoryear{{Dempsey}, {Thomas}  \& {Currie}}{{Dempsey}
  et~al.}{2013}]{2013ApJS..209....8D}
{Dempsey} J.~T.,  {Thomas} H.~S.,   {Currie} M.~J.,  2013, \mn@doi [ApJS]
  {10.1088/0067-0049/209/1/8}, \href
  {http://adsabs.harvard.edu/abs/2013ApJS..209....8D} {209, 8}

\bibitem[\protect\citeauthoryear{{Dionatos}, {Nisini}, {Codella}  \&
  {Giannini}}{{Dionatos} et~al.}{2010}]{2010A&A...523A..29D}
{Dionatos} O.,  {Nisini} B.,  {Codella} C.,   {Giannini} T.,  2010, \mn@doi
  [A\&A] {10.1051/0004-6361/200913839}, \href
  {http://adsabs.harvard.edu/abs/2010A%26A...523A..29D} {523, A29}

\bibitem[\protect\citeauthoryear{{Disney} \& {Wallace}}{{Disney} \&
  {Wallace}}{1982}]{1982QJRAS..23..485D}
{Disney} M.~J.,  {Wallace} P.~T.,  1982, QJRAS, \href
  {http://adsabs.harvard.edu/abs/1982QJRAS..23..485D} {23, 485}

\bibitem[\protect\citeauthoryear{{Draper}, {Allan}, {Berry}, {Currie},
  {Giaretta}, {Rankin}, {Gray}  \& {Taylor}}{{Draper}
  et~al.}{2005}]{2005ASPC..347...22D}
{Draper} P.~W.,  {Allan} A.,  {Berry} D.~S.,  {Currie} M.~J.,  {Giaretta} D.,
  {Rankin} S.,  {Gray} N.,   {Taylor} M.~B.,  2005, in {Shopbell} P.,
  {Britton} M.,   {Ebert} R.,  eds,  ASP Conf. Ser. Vol. 347, Astronomical Data
  Analysis Software and Systems XIV. p.~22

\bibitem[\protect\citeauthoryear{{Economou}, {Jenness}, {Tilanus}, {Hirst},
  {Adamson}, {Rippa}, {Delorey}  \& {Isaak}}{{Economou}
  et~al.}{2002}]{2002ASPC..281..488E}
{Economou} F.,  {Jenness} T.,  {Tilanus} R.~P.~J.,  {Hirst} P.,  {Adamson}
  A.~J.,  {Rippa} M.,  {Delorey} K.~K.,   {Isaak} K.~G.,  2002, in {Bohlender}
  D.~A.,  {Durand} D.,   {Handley} T.~H.,  eds,  \aspconf Vol. 281,
  Astronomical Data Analysis Software and Systems XI. p.~488

\bibitem[\protect\citeauthoryear{{Economou} et~al.,}{{Economou}
  et~al.}{2011}]{2011ASPC..442..203E}
{Economou} F.,  et~al., 2011, in {Evans} I.~N.,  {Accomazzi} A.,  {Mink} D.~J.,
    {Rots} A.~H.,  eds,  ASP Conf. Ser. Vol. 442, Astronomical Data Analysis
  Software and Systems XX. p.~203

\bibitem[\protect\citeauthoryear{Economou et~al.,}{Economou
  et~al.}{2015}]{2015Economou}
Economou F.,  et~al., 2015, \mn@doi [Astron.\ Comp.]
  {10.1016/j.ascom.2014.12.005}, 11, Part B, 161

\bibitem[\protect\citeauthoryear{{Folger}, {Bridger}, {Dent}, {Kelly},
  {Adamson}, {Economou}, {Hirst}  \& {Jenness}}{{Folger}
  et~al.}{2002}]{2002ASPC..281..453F}
{Folger} M.,  {Bridger} A.,  {Dent} B.,  {Kelly} D.,  {Adamson} A.,  {Economou}
  F.,  {Hirst} P.,   {Jenness} T.,  2002, in {Bohlender} D.~A.,  {Durand} D.,
  {Handley} T.~H.,  eds,  ASP Conf. Ser. Vol. 281, Astronomical Data Analysis
  Software and Systems XI. p.~453

\bibitem[\protect\citeauthoryear{{Graf} et~al.,}{{Graf}
  et~al.}{2003}]{2003SPIE.4855..322G}
{Graf} U.~U.,  et~al., 2003, in {Phillips} T.~G.,  {Zmuidzinas} J.,  eds,
  Proc.~SPIE Vol. 4855, Millimeter and Submillimeter Detectors for Astronomy.
  pp 322--329, \mn@doi{10.1117/12.459669}

\bibitem[\protect\citeauthoryear{{Graf}, {Honingh}, {Jacobs}, {Justen},
  {P{\"u}tz}, {Schultz}, {Wulff}  \& {Stutzki}}{{Graf}
  et~al.}{2008}]{2008stt..conf..488G}
{Graf} U.~U.,  {Honingh} C.~E.,  {Jacobs} K.,  {Justen} M.,  {P{\"u}tz} P.,
  {Schultz} M.,  {Wulff} S.,   {Stutzki} J.,  2008, in {Wild} W.,  ed.,
  Nineteenth International Symposium on Space Terahertz Technology. p.~488

\bibitem[\protect\citeauthoryear{{Graves} et~al.,}{{Graves}
  et~al.}{2010}]{2010MNRAS.409.1412G}
{Graves} S.~F.,  et~al., 2010, \mn@doi [MNRAS]
  {10.1111/j.1365-2966.2010.17140.x}, \href
  {http://adsabs.harvard.edu/abs/2010MNRAS.409.1412G} {409, 1412}

\bibitem[\protect\citeauthoryear{{Greisen}}{{Greisen}}{2003}]{2003ASSL..285..109G}
{Greisen} E.~W.,  2003, \mn@doi [Information Handling in Astronomy - Historical
  Vistas] {10.1007/0-306-48080-8_7}, \href
  {http://adsabs.harvard.edu/abs/2003ASSL..285..109G} {285, 109}

\bibitem[\protect\citeauthoryear{{Greisen}, {Calabretta}, {Valdes}  \&
  {Allen}}{{Greisen} et~al.}{2006}]{2006A&A...446..747G}
{Greisen} E.~W.,  {Calabretta} M.~R.,  {Valdes} F.~G.,   {Allen} S.~L.,  2006,
  \mn@doi [A\&A] {10.1051/0004-6361:20053818}, \href
  {http://adsabs.harvard.edu/abs/2006A%26A...446..747G} {446, 747}

\bibitem[\protect\citeauthoryear{Hatchell, Wilson, Buckle, Chrysostomou,
  Tilanus, Economou, Jenness  \& Cavanagh}{Hatchell
  et~al.}{2008}]{2008JCMTLSQA}
Hatchell J.,  Wilson C.,  Buckle J.,  Chrysostomou A.,  Tilanus R.,  Economou
  F.,  Jenness T.,   Cavanagh B.,  2008, HARP data acceptance criteria for the
  JCMT Legacy Surveys, Joint Astronomy Centre, \url
  {http://docs.eao.hawaii.edu/JCMT/JLS/QA/HARP/jls_qa_harp.pdf}

\bibitem[\protect\citeauthoryear{{Holland} et~al.}{{Holland}
  et~al.}{2013}]{2013MNRAS.430.2513H}
{Holland} W.~S.,  et~al., 2013, \mn@doi [\mnras] {10.1093/mnras/sts612}, \href
  {http://adsabs.harvard.edu/abs/2013MNRAS.430.2513H} {430, 2513}

\bibitem[\protect\citeauthoryear{{Hovey} et~al.,}{{Hovey}
  et~al.}{2000}]{2000SPIE.4015..114H}
{Hovey} G.~J.,  et~al., 2000, in {Butcher} H.~R.,  ed.,  Proc.~SPIE Vol. 4015,
  Radio Telescopes. pp 114--125, \mn@doi{10.1117/12.390404}

\bibitem[\protect\citeauthoryear{{Hurtado}, {Graf}, {Adams}, {Honingh},
  {Jacobs}, {P{\"u}tz}, {G{\"u}sten}  \& {Stutzki}}{{Hurtado}
  et~al.}{2014}]{2014SPIE.9153E..27H}
{Hurtado} N.,  {Graf} U.~U.,  {Adams} H.,  {Honingh} C.~E.,  {Jacobs} K.,
  {P{\"u}tz} P.,  {G{\"u}sten} R.,   {Stutzki} J.,  2014, in Holland W.~S.,
  Zmuidzinas J.,  eds,  \procspie\ Vol. 9153, Millimeter, Submillimeter, and
  Far-Infrared Detectors and Instrumentation for Astronomy VII. p.~27,
  \mn@doi{10.1117/12.2055563}

\bibitem[\protect\citeauthoryear{Jenness}{Jenness}{2015}]{2015HDS}
Jenness T.,  2015, \mn@doi [Astronomy \& Computing]
  {10.1016/j.ascom.2015.02.003}, in press, (\mn@eprint{arXiv}{1502.04029})

\bibitem[\protect\citeauthoryear{{Jenness} \& {Economou}}{{Jenness} \&
  {Economou}}{1999}]{1999ASPC..172..171J}
{Jenness} T.,  {Economou} F.,  1999, in {Mehringer} D.~M.,  {Plante} R.~L.,
  {Roberts} D.~A.,  eds,  \aspconf Vol. 172, Astronomical Data Analysis
  Software and Systems VIII. p.~171

\bibitem[\protect\citeauthoryear{{Jenness} \& {Economou}}{{Jenness} \&
  {Economou}}{2011}]{2011tfa..confE..42J}
{Jenness} T.,  {Economou} F.,  2011, in {Gajadhar} S.,  et~al., eds, Telescopes
  from Afar. p.~42 (\mn@eprint {} {1111.5855})

\bibitem[\protect\citeauthoryear{Jenness \& Economou}{Jenness \&
  Economou}{2015}]{2015A&C.....9...40J}
Jenness T.,  Economou F.,  2015, \mn@doi [Astronomy \& Computing]
  {10.1016/j.ascom.2014.10.005}, 9, 40

\bibitem[\protect\citeauthoryear{Jenness, Tilanus, Meyerdierks  \&
  Fairclough}{Jenness et~al.}{1999}]{GSD1999}
Jenness T.,  Tilanus R. P.~J.,  Meyerdierks H.,   Fairclough J.,  1999,
  Starlink User Note~229, The Global Section Datafile (GSD) access library.
Joint Astronomy Centre

\bibitem[\protect\citeauthoryear{Jenness, Leech, de Witt  \& Economou}{Jenness
  et~al.}{2007}]{OCS_ICD_022}
Jenness T.,  Leech J.,  de Witt S.,   Economou F.,  2007, {ACSIS File Format
  Interface Control Document}, OCS/ICD/022, Joint Astronomy Centre, \url
  {http://docs.eao.hawaii.edu/JCMT/OCS/ICD/022/ocs_icd_022.pdf}

\bibitem[\protect\citeauthoryear{{Jenness}, {Cavanagh}, {Economou}  \&
  {Berry}}{{Jenness} et~al.}{2008}]{2008ASPC..394..565J}
{Jenness} T.,  {Cavanagh} B.,  {Economou} F.,   {Berry} D.~S.,  2008, in
  {Argyle} R.~W.,  {Bunclark} P.~S.,   {Lewis} J.~R.,  eds,  ASP Conf. Ser.
  Vol. 394, Astronomical Data Analysis Software and Systems XVII. p.~565

\bibitem[\protect\citeauthoryear{Jenness et~al.,}{Jenness
  et~al.}{2014}]{2014SPIE9152-109}
Jenness T.,  et~al., 2014, in Chiozzi G.,  Radziwill N.~M.,  eds,  \procspie\
  Vol. 9152, Software and Cyberinfrastructure for Astronomy III. p. 91522W
  (\mn@eprint {arXiv} {1406.1515}), \mn@doi{10.1117/12.2056516}

\bibitem[\protect\citeauthoryear{Jenness, Stobie, Maddalena, Fairclough,
  Garwood, Prestage, Tilanus  \& Padman}{Jenness et~al.}{2015a}]{GSD2015}
Jenness T.,  Stobie E.~B.,  Maddalena R.~J.,  Fairclough J.~H.,  Garwood R.~W.,
   Prestage R.~M.,  Tilanus R. P.~J.,   Padman R.,  2015a, \mn@doi [Astronomy
  and Computing] {10.1016/j.ascom.2015.06.001}, in press, (\mn@eprint{arXiv}{1506.03136})

\bibitem[\protect\citeauthoryear{Jenness et~al.,}{Jenness
  et~al.}{2015b}]{2015NDF}
Jenness T.,  et~al., 2015b, \mn@doi [Astronomy \& Computing]
  {10.1016/j.ascom.2014.11.001}, in press, (\mn@eprint{arXiv}{1410.7513})

\bibitem[\protect\citeauthoryear{{Kloosterman} et~al.,}{{Kloosterman}
  et~al.}{2012}]{2012SPIE.8452E..04K}
{Kloosterman} J.,  et~al., 2012, in Millimeter, Submillimeter, and Far-Infrared
  Detectors and Instrumentation for Astronomy VI. p. 845204,
  \mn@doi{10.1117/12.925088}

\bibitem[\protect\citeauthoryear{{Lightfoot}, {Dent}, {Willis}  \&
  {Hovey}}{{Lightfoot} et~al.}{2000}]{2000ASPC..216..502L}
{Lightfoot} J.~F.,  {Dent} W.~R.~F.,  {Willis} A.~G.,   {Hovey} G.~J.,  2000,
  in {Manset} N.,  {Veillet} C.,   {Crabtree} D.,  eds,  ASP Conf. Ser. Vol.
  216, Astronomical Data Analysis Software and Systems IX. p.~502

\bibitem[\protect\citeauthoryear{{Maddalena}}{{Maddalena}}{2002}]{2002ASPC..278..329M}
{Maddalena} R.~J.,  2002, in {Stanimirovic} S.,  {Altschuler} D.,  {Goldsmith}
  P.,   {Salter} C.,  eds,  ASP Conf. Ser. Vol. 278, Single-Dish Radio
  Astronomy: Techniques and Applications. pp 329--352

\bibitem[\protect\citeauthoryear{{McMullin}, {Golap}  \& {Myers}}{{McMullin}
  et~al.}{2004}]{2004ASPC..314..468M}
{McMullin} J.~P.,  {Golap} K.,   {Myers} S.~T.,  2004, in {Ochsenbein} F.,
  {Allen} M.~G.,   {Egret} D.,  eds,  ASP Conf. Ser. Vol. 314, Astronomical
  Data Analysis Software and Systems (ADASS) XIII. p.~468

\bibitem[\protect\citeauthoryear{Padman}{Padman}{1990}]{1990JCMTP...9...25P}
Padman R.,  1990, PROTOSTAR: The Newsletter of the JCMT, 9, 25

\bibitem[\protect\citeauthoryear{Padman}{Padman}{1993}]{SPECX}
Padman R.,  1993, SPECX V6.3 Users' Manual.
University of Cambridge

\bibitem[\protect\citeauthoryear{{Padman} et~al.,}{{Padman}
  et~al.}{1992}]{1992IJIMW..13.1487P}
{Padman} R.,  et~al., 1992, \mn@doi [\ijimw] {10.1007/BF01009232}, \href
  {http://adsabs.harvard.edu/abs/1992IJIMW..13.1487P} {13, 1487}

\bibitem[\protect\citeauthoryear{{Petry} \& {CASA Development Team}}{{Petry} \&
  {CASA Development Team}}{2012}]{2012ASPC..461..849P}
{Petry} D.,  {CASA Development Team} 2012, in {Ballester} P.,  {Egret} D.,
  {Lorente} N.~P.~F.,  eds,  ASP Conf. Ser. Vol. 461, Astronomical Data
  Analysis Software and Systems XXI. p.~849 (\mn@eprint {arXiv} {1201.3454})

\bibitem[\protect\citeauthoryear{{Pety}}{{Pety}}{2005}]{2005sf2a.conf..721P}
{Pety} J.,  2005, in {Casoli} F.,  {Contini} T.,  {Hameury} J.~M.,   {Pagani}
  L.,  eds, SF2A-2005: Semaine de l'Astrophysique Francaise. p.~721

\bibitem[\protect\citeauthoryear{{Plume} et~al.,}{{Plume}
  et~al.}{2007}]{2007PASP..119..102P}
{Plume} R.,  et~al., 2007, \mn@doi [PASP] {10.1086/511161}, \href
  {http://adsabs.harvard.edu/abs/2007PASP..119..102P} {119, 102}

\bibitem[\protect\citeauthoryear{{Rees} et~al.,}{{Rees}
  et~al.}{2002}]{2002SPIE.4848..283R}
{Rees} N.~P.,  et~al., 2002, in {Lewis} H.,  ed.,  \procspie\ Vol. 4848,
  Advanced Telescope and Instrumentation Control Software II. pp 283--293,
  \mn@doi{10.1117/12.461306}

\bibitem[\protect\citeauthoryear{{Sadavoy} et~al.,}{{Sadavoy}
  et~al.}{2013}]{2013ApJ...767..126S}
{Sadavoy} S.~I.,  et~al., 2013, \mn@doi [\apj] {10.1088/0004-637X/767/2/126},
  \href {http://adsabs.harvard.edu/abs/2013ApJ...767..126S} {767, 126}

\bibitem[\protect\citeauthoryear{{Sandell}, {Jenness}, {McMullin}  \&
  {Shah}}{{Sandell} et~al.}{2001}]{2001AAS...19915601S}
{Sandell} G.,  {Jenness} T.,  {McMullin} J.~P.,   {Shah} R.~Y.,  2001, in
  American Astronomical Society Meeting Abstracts. p. \#156.01

\bibitem[\protect\citeauthoryear{{Schuster} et~al.,}{{Schuster}
  et~al.}{2004}]{2004A&A...423.1171S}
{Schuster} K.-F.,  et~al., 2004, \mn@doi [A\&A] {10.1051/0004-6361:20034179},
  \href {http://adsabs.harvard.edu/abs/2004A%26A...423.1171S} {423, 1171}

\bibitem[\protect\citeauthoryear{{Smith} et~al.,}{{Smith}
  et~al.}{2003}]{2003SPIE.4855..338S}
{Smith} H.,  et~al., 2003, in {Phillips} T.~G.,  {Zmuidzinas} J.,  eds,
  \procspie\ Vol. 4855, Millimeter and Submillimeter Detectors for Astronomy.
  pp 338--348, \mn@doi{10.1117/12.459674}

\bibitem[\protect\citeauthoryear{{Ward-Thompson} et~al.,}{{Ward-Thompson}
  et~al.}{2007}]{2007PASP..119..855W}
{Ward-Thompson} D.,  et~al., 2007, \mn@doi [PASP] {10.1086/521277}, \href
  {http://adsabs.harvard.edu/abs/2007PASP..119..855W} {119, 855}

\bibitem[\protect\citeauthoryear{Warren-Smith \& Berry}{Warren-Smith \&
  Berry}{2013}]{SUN211}
Warren-Smith R.~F.,  Berry D.~S.,  2013, Starlink User Note~211, AST -- A
  Library for Handling World Coordinate Systems in Astronomy.
Joint Astronomy Centre

\bibitem[\protect\citeauthoryear{{Warren} et~al.,}{{Warren}
  et~al.}{2010}]{2010ApJ...714..571W}
{Warren} B.~E.,  et~al., 2010, \mn@doi [ApJ] {10.1088/0004-637X/714/1/571},
  \href {http://adsabs.harvard.edu/abs/2010ApJ...714..571W} {714, 571}

\bibitem[\protect\citeauthoryear{{Wheeler} et~al.,}{{Wheeler}
  et~al.}{2014}]{2014SPIE.9153E..0KW}
{Wheeler} C.~H.,  et~al., 2014, in Holland W.~S.,  Zmuidzinas J.,  eds,
  \procspie\ Vol. 9153, Millimeter, Submillimeter, and Far-Infrared Detectors
  and Instrumentation for Astronomy VII. p. 91530K, \mn@doi{10.1117/12.2056606}

\bibitem[\protect\citeauthoryear{{White} et~al.,}{{White}
  et~al.}{2015}]{2015MNRAS.447.1996W}
{White} G.~J.,  et~al., 2015, \mn@doi [\mnras] {10.1093/mnras/stu2323}, \href
  {http://adsabs.harvard.edu/abs/2015MNRAS.447.1996W} {447, 1996}

\bibitem[\protect\citeauthoryear{{Whyborn}}{{Whyborn}}{1995}]{1995ASPC...75..117W}
{Whyborn} N.~D.,  1995, in {Emerson} D.~T.,  {Payne} J.~M.,  eds,  ASP Conf.
  Ser. Vol. 75, Multi-Feed Systems for Radio Telescopes. p.~117

\bibitem[\protect\citeauthoryear{{Williams}, {de Geus}  \& {Blitz}}{{Williams}
  et~al.}{1994}]{1994ApJ...428..693W}
{Williams} J.~P.,  {de Geus} E.~J.,   {Blitz} L.,  1994, \mn@doi [ApJ]
  {10.1086/174279}, \href {http://adsabs.harvard.edu/abs/1994ApJ...428..693W}
  {428, 693}

\bibitem[\protect\citeauthoryear{{Wilson} et~al.,}{{Wilson}
  et~al.}{2009}]{2009ApJ...693.1736W}
{Wilson} C.~D.,  et~al., 2009, \mn@doi [ApJ] {10.1088/0004-637X/693/2/1736},
  \href {http://adsabs.harvard.edu/abs/2009ApJ...693.1736W} {693, 1736}

\bibitem[\protect\citeauthoryear{{Young} \& {Currie}}{{Young} \&
  {Currie}}{1998}]{1998A&AS..127..367Y}
{Young} C.~K.,  {Currie} M.~J.,  1998, \mn@doi [\aaps] {10.1051/aas:1998107},
  \href {http://adsabs.harvard.edu/abs/1998A%26AS..127..367Y} {127, 367}

\makeatother
\end{thebibliography}
\end{document}